\documentclass[prd,superscriptaddress,amsfonts,amssymb,amsmath,showpacs,twocolumn]{revtex4-2}
\usepackage{bm}
\usepackage{amsfonts}
\usepackage{latexsym}
\usepackage{graphicx}
\usepackage{amsmath}
\usepackage{palatino}
\usepackage{mathpazo}
\usepackage{textcomp}
\usepackage{float}
\usepackage{booktabs}
\usepackage{dcolumn}
\usepackage{booktabs}
\usepackage{multirow}
\usepackage{hyperref}
\hypersetup{colorlinks,citecolor=blue}
\usepackage{amsmath}
\usepackage{xcolor}
\usepackage{orcidlink}
\usepackage{epsfig}
\usepackage{caption}
\usepackage{subcaption}
\usepackage{commath}
\hypersetup{colorlinks,citecolor=blue}
\hypersetup{colorlinks=true,linkcolor=magenta,filecolor=magenta,    urlcolor=blue}

\def\be{\begin{equation}}
\def\ee{\end{equation}}
\def\bea{\begin{eqnarray}}
\def\eea{\end{eqnarray}}

\begin{document}

\title{Shadows and Strong Gravitational Lensing Around Black Hole-like Compact Object in Quadratic Gravity}

\author{Somi Aktar}
\email{somiaktar9@gmail.com}
\affiliation{Department of Mathematics, Jadavpur University, Kolkata 700032, West Bengal, India}

\author{Niyaz Uddin Molla}
\email{niyazuddin182@gmail.com}
\affiliation{Department of
Mathematics, Indian Institute of Engineering Science and Technology, Shibpur, Howrah-711 103, India}

\author{Farook Rahaman}
\email{rahaman@associates.iucaa.in}
\affiliation{Department of Mathematics, Jadavpur University, Kolkata 700032, West Bengal,
India}

\author{G. Mustafa}
\email{gmustafa3828@gmail.com}
\affiliation{Department of Physics,
Zhejiang Normal University, Jinhua 321004, People’s Republic of China}


\begin{abstract}
We investigate the astrophysical consequences of black holes in quadratic gravity, characterized by the parameters $S_0$, $S_2$, $m_0$ and $m_2$, in addition to the black hole mass $M$. To evaluate the physical validity of the fundamental quadratic gravity black hole solutions, we analyze their gravitational lensing properties in the strong field regime. Specifically, we examine the shadow cast by the quadratic gravity black hole and constrain its parameters using observational data from the $M87^*$ and $Sgr A^*$ supermassive black holes. Our analysis reveals that, within the $1\sigma$ confidence level, a significant portion of the parameter space for quadratic gravity black holes is consistent with the Event Horizon Telescope (EHT) observations of $M87^*$ and $Sgr A^*$. This suggests that these black holes are plausible candidates for describing astrophysical black holes. As an additional observational test, we perform a detailed investigation of the strong gravitational lensing properties of these black holes. We explore the fundamental strong lensing observables in detail, including the angular positions and separations of the lensed images, the relative magnifications, the radius of the outermost Einstein ring and the relativistic time delay between images. We compare the predictions of the quadratic gravity black hole for each observable with those of the classical Schwarzschild solution using realistic astrophysical data. Our findings provide a pathway for testing quadratic gravity at the galactic and extragalactic scales, offering new insights into the observational properties of black hole solutions within this framework. 
\end{abstract}
\maketitle
\section{Introduction}
The study of black holes offers a unique opportunity to test the predictions of General Relativity (GR) under extreme conditions. In the high-curvature regions surrounding black holes \cite{Schwarzschild:1916uq, Kerr:1963ud}, GR predicts phenomena such as time dilation, gravitational redshift, and warping of space, which have been observed and confirmed through astrophysical measurements. The Schwarzschild and Kerr solutions to Einstein's field equations describe the spacetime around non-rotating and rotating black holes and provide the framework for understanding black hole properties. Observing black holes is crucial in gravitational and cosmological physics, unveiling the enigmatic entities that profoundly shape spacetime. 
Recent developments in observational astrophysics, such as gravitational wave detection \cite{LIGOScientific:2016aoc} from black hole mergers and direct imaging of black hole shadows \cite{EventHorizonTelescope:2022exc,EventHorizonTelescope:2019dse}, have provided empirical support for GR in strong gravitational fields. These observations not only affirm the theoretical predictions of GR, but also open new avenues for exploring potential extensions to GR, especially as researchers seek to reconcile it with quantum mechanics. Thus, black holes serve as natural laboratories, allowing scientists to probe the limits of GR and explore the fundamental nature of gravity. GR is perturbatively non-renormalizable, except for pure gravity without interactions involving scalar fields at the one-loop level \cite{Birrell:1982ix,Aharony_1999}. Consequently, GR is generally considered an effective field theory, posing significant challenges in developing an entirely consistent quantum theory of gravity. Nevertheless, gravity can be accurately described within a specific range of energy and length scales by imposing a suitable cutoff at high-energy scales. However, beyond these limits, particularly near black hole singularities, the classical formulation of GR becomes inadequate, necessitating the development of a new physical theory.

In recent years, significant advances have been made in the field of black hole physics. However, a deeper investigation of the fascinating and complex features of black holes, particularly the formation of relativistic images, is still required. Key phenomena, including gravitational lensing \cite{Kumar:2023jgh}, black hole shadows \cite{Anjum:2023axh} and quasi normal modes \cite{Cao:2023zhy}, play a crucial role in enhancing our understanding of general relativity, modified gravity theories, quantum gravity and quantum mechanics.

Despite these advances, observing such features poses considerable challenges due to technical limitations and the need for precise theoretical modeling. Increasing attention has been directed toward understanding the impact of quantum gravity effects near singularities. Models such as M-theory, string theory, and loop quantum gravity are among the leading frameworks explored in this regard. A comprehensive formulation of quantum gravity was presented in \cite{Reuter:1996cp}, where a scale-dependent action was proposed and an exact renormalization group equation was derived. This work marks a significant step forward in unraveling the intricate dynamics of quantum gravity in the vicinity of singularities.
Even when considering GR as the low-energy limit of a more comprehensive theory known as supergravity \cite{VANNIEUWENHUIZEN1981189}, the issue of non-renormalization remains unresolved. However, supergravity mitigates the ultraviolet (UV) divergences by reducing the number of divergent terms through its underlying supersymmetry. The quest for a consistent quantization of gravity encountered a deadlock until the advent of a non-perturbative approach to renormalization known as asymptotic safety \cite{Weinberg:1980gg, PhysRevD.57.971,Falkenberg:1996bq}. This framework posits the existence of a fixed point in the UV limit within the renormalization group flow \cite{Souma:1999at, PhysRevD.65.025013}, allowing the gravitational coupling constant to approach this fixed point so that physical quantities become insulated from unphysical divergences. Consequently, the resulting theory is deemed asymptotically safe gravity, implying that unphysical divergences are likely absent at high energies \cite{PhysRevD.16.953, BENEDETTI_2009}. To investigate how black holes are modified in quantum theory, one approach is to examine them within the framework of quadratic gravity. Unlike the perspective of effective field theory (EFT), quadratic curvature operators can emerge as fundamental modifications to General Relativity (GR), potentially inspired by quantum gravity principles. This perspective forms the basis for the theory of quadratic gravity \cite{hindawi1996consistent}.
It is crucial to distinguish between these two approaches. The inclusion of quadratic curvature operators introduces additional massive degrees of freedom \cite{hindawi1996consistent,stelle1978classical}. Specifically, in addition to the massless spin-2 degree-of-freedom characteristic of GR, these operators typically generate a massive spin-0 degree of freedom and a massive spin-2 degree of freedom. 
Studies of spherically symmetric black hole solutions in quadratic gravity have provided significant insights, with detailed explorations available in various works, such as\cite{bonanno2019characterizing,saueressig2021asymptotically,daas2023probing,podolsky2020black}.

In the present paper, we investigate the astrophysical scenario of black holes-like compact objects within the framework of quadratic gravity \cite{Weinberg:1980gg} alongside the quantum theory of gravity, where gravitational interactions converge toward an ultraviolet fixed point at trans-Planckian scales. Over the past three decades, this quantum field theoretical approach to gravity has emerged as a prominent contender for a consistent quantum theory of gravity, particularly in the context of the functional renormalization group (FRG) method, beginning with Reuter's seminal paper \cite{Reuter_1998}. For recent reviews on this topic,see references \cite{Reuter_Saueressig_2019,Bonanno:2020bil,Pawlowski:2020qer}.

The study of black hole shadows and gravitational lensing further enhances our understanding of GR and its predictions around these extreme objects.
The images of the supermassive black holes $M87^*$ \cite {2019ApJ...875L...4E,EventHorizonTelescope:2019dse} and $Sgr A^*$ \cite{EventHorizonTelescope:2022wkp,EventHorizonTelescope:2022apq} offer critical insights for testing gravitational effects in the strong field regime. A key characteristic of a black hole's image is its shadow, formed by light rays captured by the outer event horizon. The shape and size of the shadow depend on several factors, including the parameters of the black hole, the paths followed by the light rays, and the observer's location. When light rays approach a black hole, photons with low orbital angular momentum are trapped by the black hole, while those with higher angular momentum escape their gravitational pull. As a result, a distant observer perceives a dark region in the sky, known as the black hole's shadow. This shadow, a silhouette of the event horizon outlined against the bright background of accreting material, is a direct observational indicator of the black hole's presence and characteristics. The shadow shape and size depend on the black hole's mass, spin, and curvature of the surrounding spacetime, offering a valuable opportunity to test the predictions of GR in the strong-field regime.
For a detailed discussion of black hole shadows and their potential for testing GR, see \cite{PhysRevD.100.044057,PhysRevD.100.024020,Kumar_2020,PhysRevD.101.084041,PhysRevD.103.025005,Bronzwaer:2021lzo,SINGH2022168892,MOLLA2023169304,PhysRevD.105.083002,ANJUM2023101195,Afrin_2021,Hamil_2023,Perlick_2022,Mirzaev:2023oud,Narzilloev:2021ifl,Al-Badawi:2024dzc,Ror:2024vgh,_vg_n_2023,_vg_n_2024,Contreras_2020,Aliyan_2024,Nozari_2023,konoplya2024probingeffectivequantumgravity}.
\\

Gravitational lensing has become an invaluable tool for probing fundamental properties of gravitational fields. It plays a crucial role in the study of the large-scale structure of the Universe, the behavior of massive stellar objects and the search for dark matter candidates. In this work, we suggest that gravitational lensing can also be employed to distinguish between different gravitational theories, particularly between asymptotically safe quantum gravity and general relativity. 
The deflection of light by black holes and galaxies and the phenomenon of strong gravitational lensing form a significant area of research. It can potentially reveal novel astrophysical effects, providing unique signatures for testing modified theories of gravity. Observables obtained from relativistic images in strong lensing scenarios could help distinguish between different black hole metrics through high-precision observations.
Gravitational lensing, a phenomenon predicted by Einstein in 1915 \cite{Einstein:1936llh}, although he initially doubted it would ever be observed. This occurs when massive objects like black holes bend and magnify light from objects behind them due to their immense gravitational fields. In the black hole regime, gravitational lensing provides a unique way to study the properties of these enigmatic objects and their influence on spacetime \cite{1992grle.book.....S,Virbhadra:1999nm,Virbhadra:1998dy,Virbhadra:2022iiy,Virbhadra:2008ws,Virbhadra:2022iiy,Virbhadra:2022ybp,Virbhadra:2007kw}. When light from a distant source, such as a star or galaxy, passes close to a black hole, it can be bent into arcs or multiple images, depending on the alignment. This effect also allows us to observe a black hole’s shadow as the intense gravity bends light around the black hole, casting a silhouette against the backdrop of surrounding radiation. This effect not only magnifies distant objects but also distorts and duplicates their images, revealing information about the black hole’s mass and spin and the geometry of spacetime near the event horizon. Recent images from the Event Horizon Telescope (EHT), capturing the shadow of the supermassive black hole in the galaxy $M87$, have provided empirical evidence that aligns closely with GR’s theoretical predictions, showcasing both the accuracy and limitations of the theory. These observations contribute to our understanding of the gravitational dynamics around black holes while presenting new challenges and opportunities for testing GR under conditions inaccessible in any other environment. There are three primary types of gravitational lensing, classified based on the alignment of the source, lens and observer, as well as the resulting effects on the observed light, each offering unique insights into the nature of these extreme objects. On the other hand, weak gravitational lensing involves minor distortions in the shape of background objects when light passes through a less intense gravitational field near a black hole. Microlensing occurs when a black hole or other compact object momentarily magnifies the light of a background star without creating multiple images. Strong gravitational lensing is particularly significant in the study of black holes because it produces highly visible and distinct effects, such as multiple images, arcs, or Einstein rings, which can be directly observed and analyzed. The clarity and magnitude of the distortions caused by strong lensing make it an exceptionally powerful tool for probing the fundamental nature of black holes and the behavior of light and matter in their extreme gravitational fields. The study of strong gravitational lensing around black holes has made substantial progress since Darwin's initial work on photon orbits \cite{Darwin:1959md}, in the Schwarzschild black hole spacetime.
Building on this foundation, Virbhadra and Ellis formalized the concept of relativistic images \cite{Virbhadra:1999nm}, focusing on the strong-field regime where light passing near a black hole's photon sphere can generate multiple, highly curved images. A significant breakthrough came with Bozza's development of the strong deflection limit (SDL) \cite{Bozza:2002zj}, which provided an accurate framework for calculating light deflection in the proximity of black holes. Further advancements were made by Frittelli, Killing and Newman, who enhanced analytical methods for solving the lens equation \cite{Frittelli:1999yf}, while Bozza extended the SDL approach. This approach has been utilized across various spacetimes, including rotating and charged black holes, such as Reissner-Nordström and Kerr black holes  \cite{Eiroa:2002mk,Bozza:2002af}. The SDL has proven to be a versatile tool, applied to classical black hole spacetimes and to modified scenarios like braneworld black holes and those arising in modified gravity theories \cite{ PhysRevD.69.063004, Whisker:2004gq, Eiroa:2012fb, Bhadra:2003zs,Kumar:2022fqo,Kumar:2021cyl,Islam:2021dyk}. Ongoing research continues to investigate gravitational lensing in diverse black hole spacetimes, including those described by higher-curvature gravity theories \cite{Kumar:2020sag, Islam:2020xmy, Narzilloev:2021jtg,Islam:2022ybr,Kumar:2020sag}, as well as modifications of the traditional Schwarzschild geometry \cite{Eiroa:2010wm, Ovgun:2019wej, Panpanich:2019mll, Bronnikov:2018nub,Molla:2024lpt}. 
Several researchers have extensively explored the physical observables associated with strong lensing effects across a variety of gravitational models and black hole solutions. These studies have analyzed various astrophysical implications by examining parameters such as the angular positions of images, separations between them, magnifications, Einstein rings and time delays in the formation of relativistic images have been explored extensively in \cite{Islam:2021ful,Kumar:2021cyl,Islam:2022ybr,Kumar:2020hgm,Molla:2022mjl,Molla:2023hou,Narzilloev:2021jtg}. The current study focuses on the shadow and strong gravitational lensing features of black holes-like  solution in quadratic gravity, highlighting distinctive geometric properties that set them apart from other well-known black hole solutions.
Notable examples  include the Kerr black hole \cite{Cunha:2018acu}, the Reissner-Nordström (RN) black hole \cite{Pang:2018jpm}, the Van der Waals black hole \cite{Molla:2022izk}, the Simpson-Visser black hole \cite{Islam_2021}, the acoustic charged black hole \cite{Molla:2023hou}, the hairy Kerr black hole \cite{Afrin_2021}, the Kerr-Taub-NUT spacetime \cite{Shao-Wen_Wei_2012}, the Kerr-Taub-NUT-Quintessence black hole \cite{Molla:2021sgw}, the Bardeen black hole \cite{Molla:2024sig}, and black holes arising from Horndeski gravity \cite{Bessa:2023ykd}, among others.

Gravitational lensing effects caused by these black holes have been analyzed both numerically and analytically in the strong-field regime \cite{Zhao:2016kft,Qiao:2022nic,Feleppa:2024vdk}. Kumar et al. \cite{Kumar:2021cyl} investigated the strong gravitational lensing of a hairy black hole in Horndeski gravity, discussing its astrophysical implications for supermassive black holes and comparing the results with the Schwarzschild black hole in General Relativity. Similarly, Islam et al. \cite{Islam:2021ful} examined the strong gravitational lensing of the rotating Simpson-Visser black hole, demonstrating its astrophysical significance and its distinct properties compared to the Kerr black hole. Islam et al. \cite{Islam:2022ybr} analyzed the strong gravitational lensing properties of the Bardeen black hole in the context of four-dimensional Einstein-Gauss-Bonnet (EGB) gravity. They examined the astrophysical implications by modeling supermassive black holes with the 4D EGB Bardeen black hole and compared their findings to those of standard Schwarzschild and Bardeen black holes. Similarly, Kumar et al. \cite{Kumar:2022fqo} investigated gravitational lensing in the strong-field regime for regular electrically charged (REC) black hole spacetimes, as well as REC no-horizon spacetimes. Their study highlighted the astrophysical significance of supermassive black holes described by regular spacetimes, with comparisons drawn to the Schwarzschild black hole.

This paper aims to extend the analysis of black hole shadows and strong gravitational lensing by incorporating the spacetime geometry of black hole-like solutions in quadratic gravity. We explore how this promising framework, rooted in the black hole-like solutions of quadratic gravity, leads to potentially observable consequences for black hole spacetimes. In particular, we focus on key lensing observables, such as the black hole shadow, deflection angle, image magnifications and time delays, and compare them with predictions from the classical lensing scenario around a black hole in a vacuum.
By analyzing the interaction between the gravitational and quantum fields of the black hole, we aim to constrain the parameters of the black hole through observational signatures. These results carry significant implications for understanding the environment surrounding black holes, testing GR and alternative gravity theories, and probing the fundamental properties of quadratic gravity through astrophysical observations.

The present paper is organized as follows: Section~\ref{sect1} provides a brief overview of the black hole solutions in asymptotically safe quantum gravity, with a focus on the non-rotating case. In Section~\ref{sect2}, we derive the expression for the black hole shadow in quadratic gravity and present the first set of constraints on the solution parameters based on comparisons with observational data from $M87^*$ and $Sgr A^*$. Section~\ref{sect3} explores the strong gravitational lensing properties of the quadratic gravity black hole, with a detailed investigation of the associated lensing observables. Specific predictions for these observables are made using realistic astrophysical data. Finally, the conclusions and implications of our findings are discussed in Section~\ref{sect4}.

In what follows, we use natural units with G = c = 1 and adopt a signature convention of \{- + + +\}\\\\

\section{Black hole-like solutions in quadratic gravity}\label{sect1}
Quadratic gravity plays a key role in modern studies of relativistic quantum field theories. As a natural extension of Einstein’s theory, the most precise gravity framework to date, quadratic gravity introduces quadratic terms in the Lagrangian that act as corrections to general relativity, becoming significant at extremely high energies. These higher-order curvature correctios are crucial for developing a consistent quantum gravity theory, applicable near the Big Bang or spacetime singularities within black holes. Remarkably, as shown by Stelle in the 1970s \cite{stelle1977renormalization} and reaffirmed in the literature \cite{salvio2018quadratic,podolsky2020black}, the inclusion of quadratic curvature terms renders gravity renormalizable, even when coupled with a generic quantum field theory.

To explore how black holes are modified within the framework of quantum theory, one can investigate them in the context of quadratic gravity.
 Black holes in quadratic gravity are described by solutions derived from the quantum effective action and the associated quantum equations of motion. These solutions are formulated using multi-graviton correlation functions within quadratic gravity framework and have been recently explored numerically by Pawlowski and Tränkle \cite{Pawlowski:2023dda}.
 The effective action is given by  
\begin{equation}\label{ee1}
\begin{split}
&\Gamma[g_{\mu\nu}] = \frac{1}{16\pi} 
\int_x d^4x \, \sqrt{g} 
\bigg[ R(\Delta, R) 
+ R \, f_{R^2}(\Delta) \, R \\
&~~~~~~+ R_{\mu\nu} \, f_{R_{\mu\nu}^2}(\Delta) \, R^{\mu\nu} 
\bigg].\\ 
\end{split}
\end{equation}
Here, the functions $f_{R^2}(\Delta)$ and $f_{R_{\mu\nu}^2}(\Delta)$ correspond to two-point correlation functions of the curvature scalar and Ricci tensor, respectively. These functions, often referred to as form factors, capture generalized dispersion effects. Studies of such form factors have been presented in works such as \cite{denz2018towards} and \cite{knorr2021nonperturbativepropagatorsquantumgravity,knorr2021formfactorsquantumgravity,Draper_2020,platania2023divergingblackholeentropy}. For a comprehensive review of recent developments, see \cite{knorr2024form}.
To derive black hole-like solutions within the framework in quadratic gravity, one must solve the field equations obtained from the effective action $\Gamma[\bar{g}]$ given by Eq.~\eqref{ee1} \cite{Pawlowski:2023dda,stelle1978classical}.

The static, spherically symmetric metric in quadratic gravity can be expressed as \cite{stelle1978classical,Pawlowski:2023dda}:
\begin{equation} \label{e1}
ds^2 = -f(r) dt^2 + g(r)^{-1} dr^2 + r^2 ( d\theta^2 + \sin^2 \theta d\phi^2),
\end{equation}
where the metric function can be expressed as:
\begin{equation}\label{e2}
f(r ) = 1 - \frac{2M}{r} + S_0 \frac{e^{-m_0 r}}{r} + S_2 \frac{e^{-m_2 r}}{r},
\end{equation}

\begin{equation}\label{e3}
g(r ) = 1 - \frac{2M}{r} - S_0 \frac{e^{-m_0 r}}{r} (1 + m_0 r) 
+ \frac{1}{2} S_2 \frac{e^{-m_2 r}}{r} (1 + m_2 r). 
\end{equation}
The two free parameters $S_0$ and $S_2$ govern the strength of the exponentially decaying Yukawa corrections. The masses $m_0$ and $m_2$ correspond to the spin-0 and spin-2 modes, respectively and are linked to the couplings of various curvature terms. For the initial conditions $(S_0, S_2) = (0, 0)$, the asymptotically safe quantum gravity black hole metric Eq.\eqref{e1} reduces to the classical Schwarzschild metric. The event horizon, located at $r_h$, is determined by the solution to the equation $g(r_h) = 0$.

\section{Shadow of a black hole in quadratic gravity} \label{sect2}
In this section, we begin our investigation of the black hole shadow within the quadratic gravity framework, focusing on the motion of photons and the resulting shadow. Our analysis begins by examining the trajectories of light in the space-time of a quadratic gravity black hole. Furthermore, we aim to constrain the black hole parameters associated with this quadratic gravity model.
Now using Eq.~(\ref{e2}) one can consider the function $\gamma(r)$  \cite{Perlick:2021aok,Ghorani:2023hkm,vagnozzi2023horizon}:
\begin{eqnarray}\label{e4}
\gamma^2(r)=\frac{r^2}{f(r)}.
\end{eqnarray}

The radius of a circular light orbit, particularly the one defining the photon sphere with radius $r_{ph}$, is found by solving the following condition \cite{Perlick:2021aok}:  
\begin{equation}\label{e5}
\frac{d(\gamma^2(r))}{dr}\bigg|_{r=r_{{ph}}} = 0. 
\end{equation}  
The photon orbit radii $r_{ph}$ in quadratic gravity can be determined numerically using this equation.  

When the parameters are $S_0 = S_2 = 0$, the solution yields $r_{ph} = 3M$ \cite{Perlick:2021aok,vagnozzi2023horizon}, corresponding to the radius of the photon orbit in the Schwarzschild spacetime.

Evaluating Eq.~\eqref{e5} for the linearized solution, as outlined in the literature \cite{daas2023probing}, the photon sphere radius $r^L_{ph}$ in the linear approximation for the metric \eqref{e1} is given by:
\begin{equation}
r^L_{ph} = 3M + \frac{3}{2} S_{2} (m_2 M - 1) e^{-3m_2 M} + \frac{3}{2}S_{0} (m_0 M - 1) e^{-3m_0 M}.
\end{equation}  

\begin{figure*}[htbp]
 \captionsetup[subfigure]{labelformat=simple}
    \renewcommand{\thesubfigure}{(\alph{subfigure})}
		\begin{subfigure}{.45\textwidth}
			\caption{}\label{sn_01a}
			\includegraphics[height=3in, width=3.in]{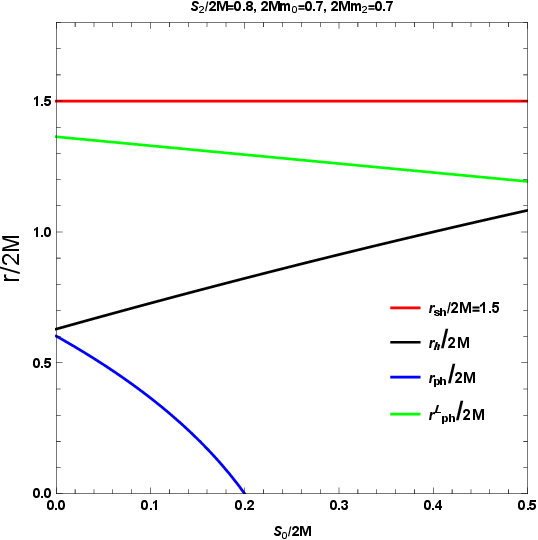}
		\end{subfigure}
            \begin{subfigure}{.4\textwidth}
			\caption{}\label{sn_01b}
			\includegraphics[height=3in, width=3.in]{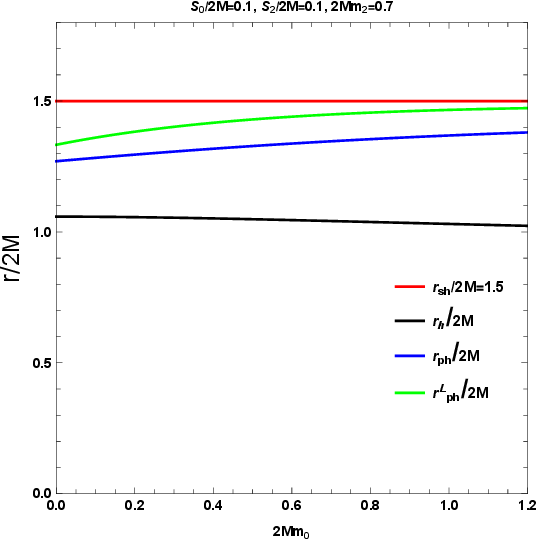}
		\end{subfigure}
		\caption{Behaviour of event horizon radius $r_h/2M$, photon sphere radius for standard Schwarzschild black hole $r_{sh}/2M=1.5$, standard photon sphere radius $r_{ph}/2M$ and new photon sphere radius $r^L_{ph}/2M$ with the parameters $S_0/2M$ (left panel) and $2M m_0$ (right panel), keeping the other parameters fixed.}
		\label{fsn01}
\end{figure*} 

\begin{table*}[htbp]

\begin{tabular}{| p{1.5cm} | p{1.5cm} | p{1.cm} | p{1.4cm} | p{1.cm} | p{2cm} | p{1.6cm} | p{1.8cm} | p{1.6cm} |}
\hline
 $S_0/2M$ & $2Mm_0$ &$S_2/2M$ & $2Mm_2$ &$r_{h}/2M$&$r_{ph}/2M$& $r^L_{ph}/2M$&$R_{sh}/2M$& $R^L_{sh}/2M$\\
 \hline
~0&~~~ 0& 0 &~~  &~~~~~~1&1.5&1.5&2.598&2.598\\
\hline

 &  &  & $0.3$ & 1.042 & 1.289 &1.406&2.287&2.31\\
 &  & $0.1$ & $0.6$ &1.047 & 1.32& 1.426&2.349&2.367\\
 &  &  & $0.9$ & 1.052 & 1.343 & 1.436&2.39&2.404\\
 &  &  & 1.1&1.056&1.356 &1.44&2.41&2.422\\
 &  &  &  & & & &&\\
 &  &  & && &&&\\
 &  &  & $0.3$ &0.995& 1.164 & 1.366&2.09&2.154\\
 & $0.5$ & $0.2$ & $0.6$ & 1.003 & 1.223 & 1.404&2.214&2.265\\
 &  &  & $0.9$ &1.014&1.272 &1.425&2.301&2.338\\
 &  &  & 1.1&1.022&1.299 &1.434&2.343&2.372\\
 &  &  & && &&&\\
  &  &  &  & &&& &\\
 &  &  & $0.3$ & 0.947 &1.031 &1.325&1.877&2.006\\
 &  & $0.3$ & $0.6$ & 0.959& 1.116& 1.383&2.062&2.166\\
 &  &  & $0.9$ &0.974 &1.192 &1.415&2.201&2.273\\
 &  &  & 1.1&985& 1.234&1.427&2.268&2.323\\
 &  &  & && &&&\\
$0.1$ &  &  &  & & &&&\\
&  &  & && &&&\\
 &  &  & $0.3$ &1.025 & 1.329 &1.443&2.357&2.378\\
 &  & $0.1$ & $0.6$ &1.029 &1.36 &1.462&2.419&2.436\\
 &  &  & $0.9$ &1.034 &1384 &1.473&2.461&2.473\\
 &  &  & 1.1&1.038& 1.397&1.476&2.48&2.49\\
 &  &  & && &&&\\
 &  &  &  & &&& &\\
 &  &  & $0.3$ & 0.978&1.203 &1.402&2.162&2.222\\
 & $1$ & $0.2$ & $0.6$ &0.986 & 1.264&1.441&2.288&2.335\\
 &  &  & $0.9$ & 0.996& 1.314&1.462&2.375&2.408\\
 &  &  & 1.1&1.004& 1.341&1.47&2.417&2.442\\
 &  &  &&& &&&\\
  &  &  &  & & &&&\\
 &  &  & $0.3$ &0.931 &1.069 &1.361&1.951&2.074\\
 &  & $0.3$ & $0.6$ &0.942 &1.158 &1.419&2.14&2.237\\
 &  &  & $0.9$ & 0.957 &1.126 &1.451&2.279&2.344\\
 &  &  & 1.1&0.968&1.278&1.464&2.346&2.395\\
 
\hline
\end{tabular}
\caption{ Estimation of the event horizon radius $r_h/2M$, standard photon sphere radius $r_{ph}/2M$, new photon sphere radius $r^L_{ph}/2M$,  black hole shadow radius $R_{sh}/2M$ corresponding to $r_{ph}/2M$ and black hole shadow radius $R^L_{sh}/2M$ corresponding to $r^L_{ph}/2M$ for different values of the  parameters  $S_0=0,0.1$,  $S_2=0,0.1,0.2,0.3$, $ m_0=0.5,1$ and $ m_2=0.3,0.6,0.9,1.1$ respectively.}\label{table:01}
\end{table*}
From the perspective of quadratic gravity, the geometry described by Eqs. \eqref{e1}-\eqref{e3} represents the gravitational field in the linear approximation. However, this framework breaks down in the strong gravity regime, where the event horizon becomes unstable due to the inclusion of exponentially small Yukawa corrections. Additionally, non-zero values of $S_0$ and $S_2$ can give rise to wormholes and naked singularities, leading to significant deviations from standard black hole geometry. These deviations necessitate careful consideration when applying the standard formulas for shadow imaging, as new photon paths may emerge that are otherwise blocked by the event horizon in conventional scenarios. To address these complexities, we have calculated the event horizon radius $r_h$, the standard photon sphere radius $r_\text{ph}$ and the new photon sphere radius $r_\text{ph}^L$, obtained through a linearized solution approximation. Our analysis shows that $r_h < r_\text{ph} < r_\text{ph}^L$ (see Table \ref{table:01}) based on our specific choice of parameter set, indicating that the photon sphere and its modifications, such as the new photon sphere obtained through the linearized solution approximation, play a crucial role in determining the shadow size. From Fig. \ref{fsn01}, it is observed that for an arbitrarily chosen parameter set, the relation $r_\text{ph} < r_\text{ph}^L$ is maintained. However, the standard photon sphere radius can sometimes be smaller than the event horizon radius depending on the parameter choice, implying that the standard photon sphere may be blocked by the event horizon in certain cases. Otherwise, when both the standard photon sphere and the new photon sphere are present, this dual-photon-sphere configuration significantly influences the formation of the black hole shadow.

We now analyze the shadow of the black hole within the quadratic gravity framework.  

The angular radius of the black hole shadow is given by the following relation \cite{Perlick:2021aok}:  
\begin{eqnarray}\label{shadow nonrotating1}
\sin^2 \alpha_{sh} = \frac{\gamma^2(r_{ph})}{\gamma^2(r_{obs})},
\end{eqnarray}  
where $\alpha_{sh}$ denotes the angular radius of the black hole shadow and $r_{obs}$ is the observer’s distance, assumed to be very large but finite. For instance, the observer distance for $Sgr A^*$ is $r_{obs} = D \simeq 8.3 \, \text{kpc}$ \cite{EventHorizonTelescope:2022wkp}, while for $M87^*$, it is $r_{obs} = D \simeq 16.8 \, \text{Mpc}$ \cite{EventHorizonTelescope:2019dse}.
 
 The quantity $r_{ph}$ represents the radius of the photon sphere, as mentioned earlier. By combining Eqs.~(\ref{e4}) and ~(\ref{e5}), it follows that for a distant observer, Eq.~(\ref{shadow nonrotating1}) can be rewritten as \cite{Perlick:2021aok,2023ChPhC..47b5102A}:
\begin{eqnarray}\label{e6}
\sin^2 \alpha_{sh} = \frac{r_{ph}^2}{f(r_{ph})} \frac{f(r_{obs})}{r_{obs}^2}.
\end{eqnarray}
The radius of the black hole shadow for an observer at large distances can be determined using Eq.~(\ref{e6}), yielding the following expression \cite{Perlick:2021aok,2023ChPhC..47b5102A}:
\begin{eqnarray}\label{shadow nonrotating3}
R_{sh} &\simeq& r_{obs} \sin \alpha_{sh} \simeq \frac{r_{ph}}{\sqrt{f(r_{ph})}}.
\end{eqnarray}
This equation allows us to examine how the shadow radius of the black hole depends on the parameters of quadratic gravity.

\begin{figure*}[htbp]
 \captionsetup[subfigure]{labelformat=simple}
    \renewcommand{\thesubfigure}{(\alph{subfigure})}
		\begin{subfigure}{.45\textwidth}
			\caption{}\label{sn1a}
			\includegraphics[height=3in, width=3in]{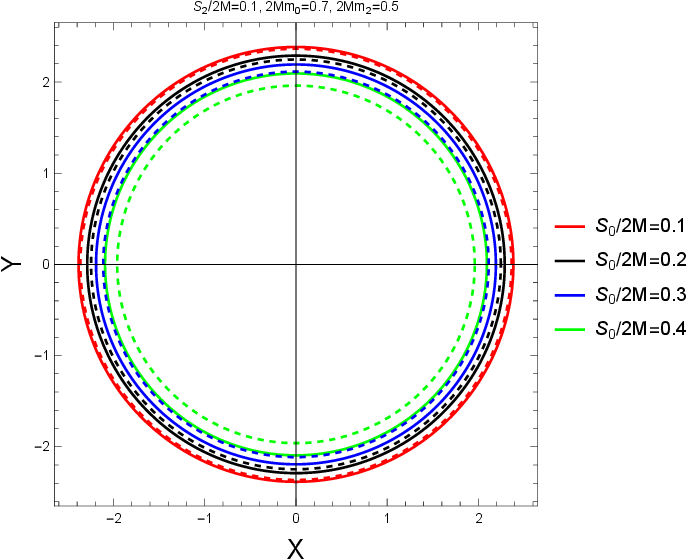}
		\end{subfigure}
            \begin{subfigure}{.4\textwidth}
			\caption{}\label{sn1b}
			\includegraphics[height=3in, width=3in]{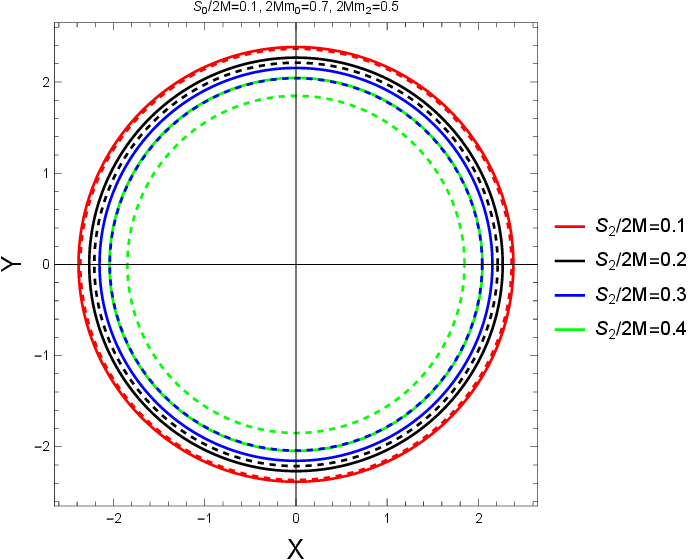}
		\end{subfigure}
  \begin{subfigure}{.45\textwidth}
			\caption{}\label{sn1c}
			\includegraphics[height=3in, width=3in]{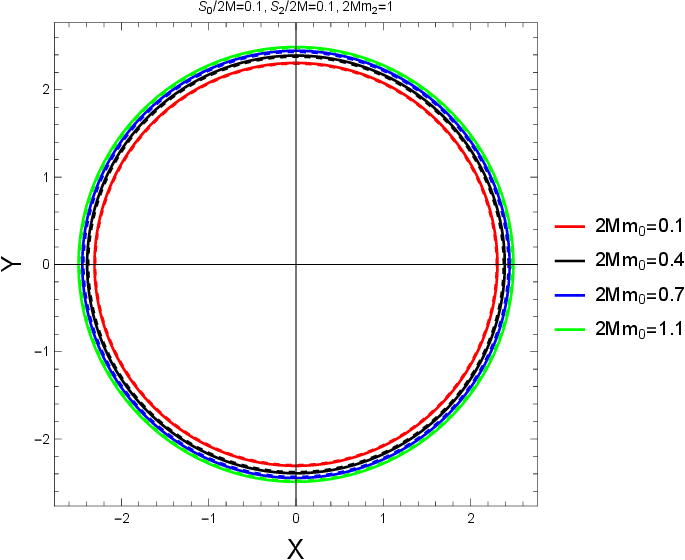}
		\end{subfigure}
            \begin{subfigure}{.4\textwidth}
			\caption{}\label{sn1d}
			\includegraphics[height=3in, width=3in]{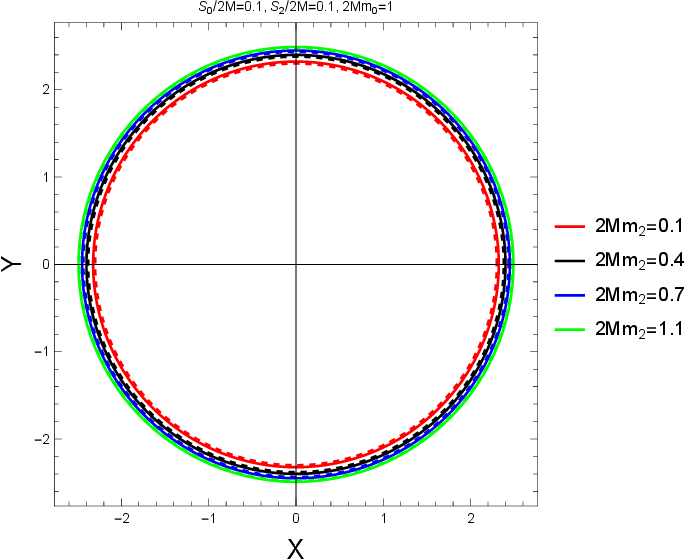}
		\end{subfigure}
		\caption{The shadow silhouette of the black hole in quadratic gravity is presented in the X-Y plane ($R_{sh}=\sqrt{X^2+Y^2}$ ), with black hole parameters $S_0$,$S_2$,$m_0$ and $m_2$ varied independently, while the other parameters are kept fixed. The dotted circular ring correspondence to shadow radius $R_{sh}/2M$ corresponding to $r_{ph}/2M$ and the solid circular ring correspondence to shadow radius $R^L_{sh}/2M$ corresponding to new photon sphere radius $r^L_{ph}/2M$ $r^L_{ph}/2M$. }
		\label{fsn1}
\end{figure*}

The dependence of the black hole shadow on the parameters $S_0,~S_2,~m_0$ and $m_2$ is illustrated in Fig.~\ref{fsn1}. It is evident that as the numerical values of the parameters $S_0 > 0$ or $S_2 > 0$ increase while the other parameters remain fixed, the size of the black hole shadow decreases. Conversely, an increase in the values of the parameters $m_0 > 0$ or $m_2 > 0$, with the other parameters held constant, increases the size of the black hole shadow.
We find that the shadow size calculated using the standard photon sphere radius $r_\text{ph}$ is smaller than that obtained using the new photon sphere radius $r_\text{ph}^L$. These results underscore the importance of incorporating the effects of the new photon sphere radius when investigating shadow imaging in quadratic gravity, particularly in the strong gravity regime.

\subsection{Constraints the black hole  parameters with the EHT observational data from $M87^*$ and $Sgr A^*$}

Here, we aim to constrain the possible values of the free parameters of black holes in quadratic gravity by analyzing observational data from supermassive black holes $M87^*$ and $Sgr A^*$. Specifically, we will explore the theoretical constraints on the black hole parameters $S_0$, $S_2$, $m_0$ and $m_2$ pairwise while keeping the remaining two parameters fixed. 

To achieve this, we use the shadow data provided by the Event Horizon Telescope (EHT) project. The objective is to constrain the parameter pairs $(S_0, S_2)$ and $(m_0, m_2)$ by fitting them against the observed shadow sizes of $M87^*$ and $Sgr A^*$ within the framework of quadratic gravity. This approach will allow us to impose meaningful constraints on these parameters using real observational data.

The angular diameter of the shadow, the distance from the Sun to the black hole, and the mass of the black hole at the center of the galaxy M87 are measured to be $\Omega_{\text{M87*}} = 42 \pm 3 \, \mu as$, $D = 16.8 \pm 0.8 \, \text{Mpc}$, and $M_{\text{M87*}} = (6.5 \pm 0.7) \times 10^9 \, M_\odot$, respectively \cite{Akiyama_2019}.

For $Sgr A^*$, the most recent data obtained by the EHT project are $\Omega_\text{Sgr A*} = 51.8 \pm 2.3 \:\mu as$, $D = 8277 \pm 9 \pm 33 \:\text{pc}$ and $M_\text{Sgr A*} = 4.297 \pm 0.013 \times 10^6 \: M_\odot$ (from VLTI observations) \cite{EventHorizonTelescope:2022wkp}. 

Based on these observational data, one can estimate the diameter of the black hole's shadow per unit mass using the following expression\cite{2023ChPhC..47b5102A,Arora:2023ijd, Bambi:2019tjh}:
\begin{equation}
    d_\text{sh} = \frac{D \, \Omega}{M} \,.
    \label{43}
\end{equation}

The diameter of the shadow can be determined using the formula $d_{\text{sh}}^{\text{theo}} = 2R_{\text{sh}}$. Thus, the diameter of the black hole shadow is $d_{\text{sh}}^{\text{Sgr A*}} = (9.5 \pm 1.4)M$ for $Sgr A^*$ and $d_{\text{sh}}^{\text{M87*}} = (11 \pm 1.5)M$ for $M87^*$ . 

\begin{figure*}[htbp]
 \captionsetup[subfigure]{labelformat=simple}
    \renewcommand{\thesubfigure}{(\alph{subfigure})}
		\begin{subfigure}{.45\textwidth}
			\caption{}\label{sn2a}
			\includegraphics[height=3in, width=3in]{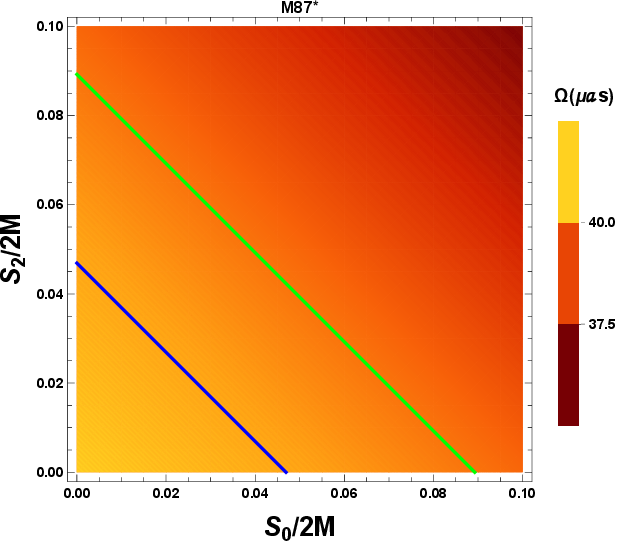}
		\end{subfigure}
            \begin{subfigure}{.4\textwidth}
			\caption{}\label{sn2b}
			\includegraphics[height=3in, width=3in]{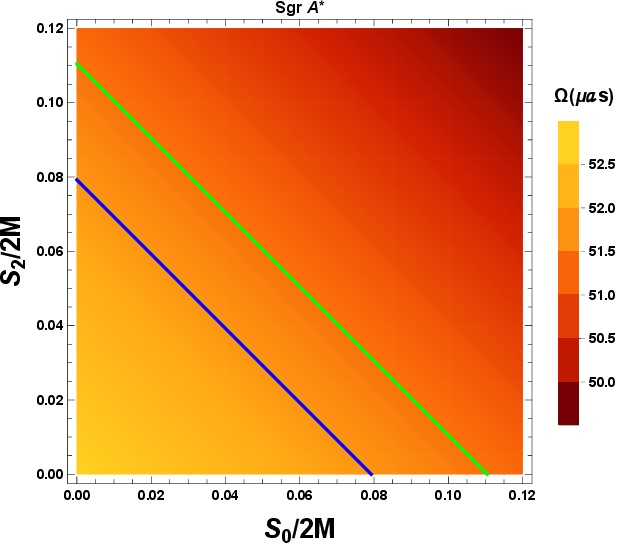}
		\end{subfigure}
  \begin{subfigure}{.45\textwidth}
			\caption{}\label{sn2c}
			\includegraphics[height=3in, width=3in]{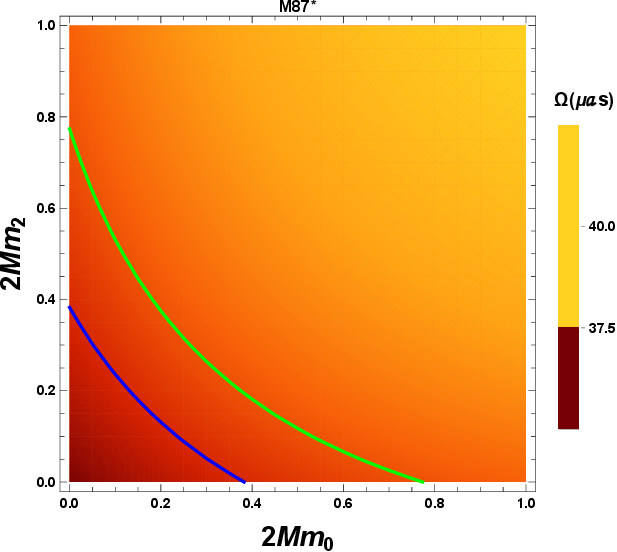}
		\end{subfigure}
            \begin{subfigure}{.4\textwidth}
			\caption{}\label{sn2d}
			\includegraphics[height=3in, width=3in]{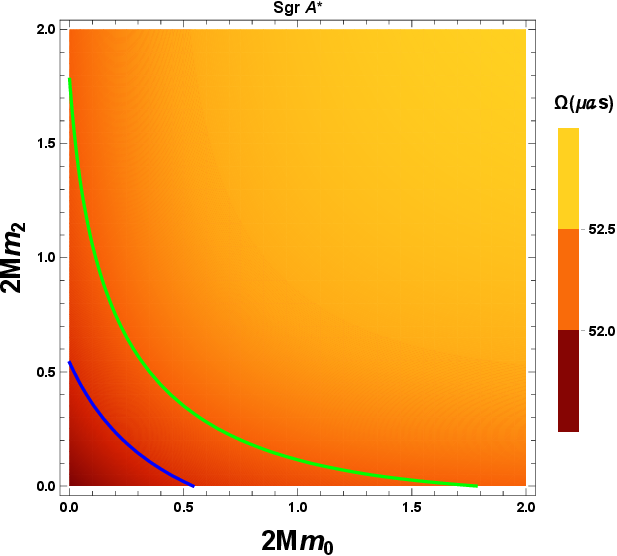}
		\end{subfigure}
		\caption{The angular diameter $\Omega(\mu as)$ of the black hole shadow correspondence to standard photon sphere radius $r_{ph}/2M$ is presented as a function of the parameters  $S_0,~ S_2$ and  $m_0, ~m_2$ for the $M87^*$ (panels (a) and (c)) and $Sgr A^*$ (panels (b) and (d)) supermassive black holes. The enclosed regions satisfy the 1-\(\sigma\) shadow bounds for $M87^*$ and $Sgr A^*$, respectively. In panel (a), the green and blue lines correspond to $\Omega= 39.1 \, \mu as$ and $\Omega= 39.5 \, \mu as$, respectively; in panel (b), they correspond to $\Omega = 51.4 \, \mu as$ and $\Omega = 51.8 \, \mu as$, respectively; in panel (c), to $\Omega = 39.4 \, \mu as$ and $\Omega = 39.3 \, \mu as$, respectively; and in panel (d), to $\Omega = 52.2 \, \mu as$ and $\Omega = 52 \, \mu as$, respectively.}
\label{fsn2}
\end{figure*}

\begin{figure*}[htbp]
 \captionsetup[subfigure]{labelformat=simple}
    \renewcommand{\thesubfigure}{(\alph{subfigure})}
		\begin{subfigure}{.45\textwidth}
			\caption{}\label{sn02a}
			\includegraphics[height=3in, width=3in]{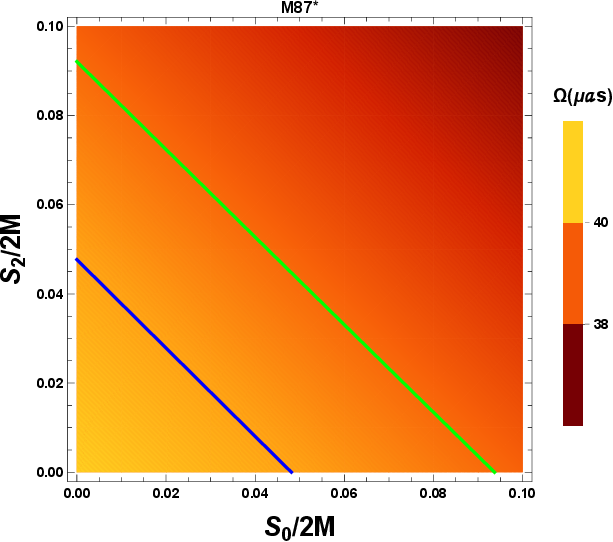}
		\end{subfigure}
            \begin{subfigure}{.4\textwidth}
			\caption{}\label{sn02b}
			\includegraphics[height=3in, width=3in]{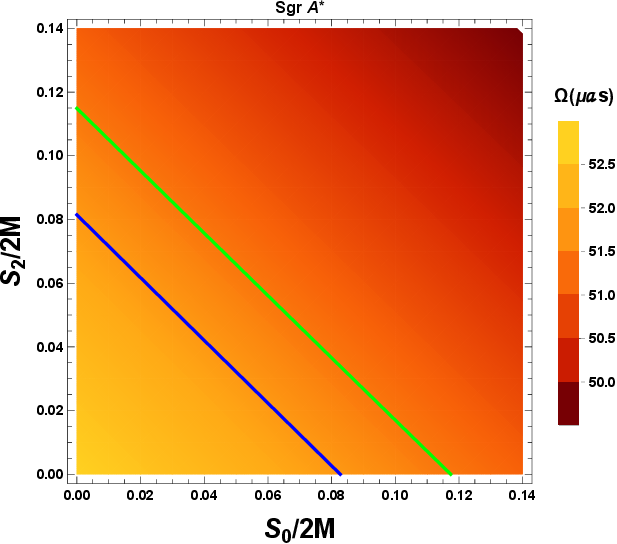}
		\end{subfigure}
  \begin{subfigure}{.45\textwidth}
			\caption{}\label{sn02c}
			\includegraphics[height=3in, width=3in]{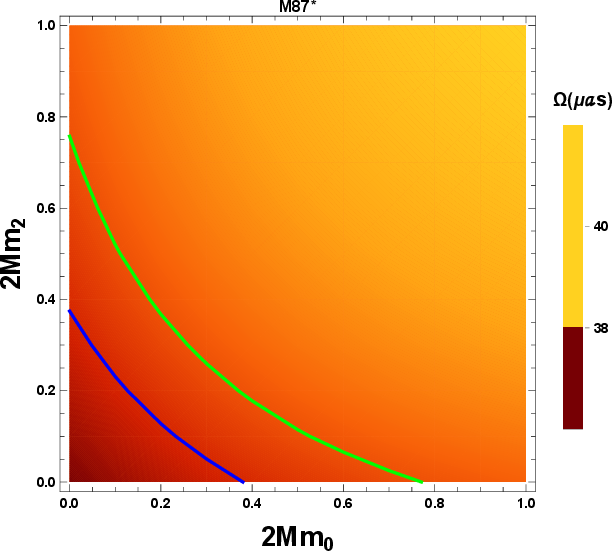}
		\end{subfigure}
            \begin{subfigure}{.4\textwidth}
			\caption{}\label{sn02d}
			\includegraphics[height=3in, width=3in]{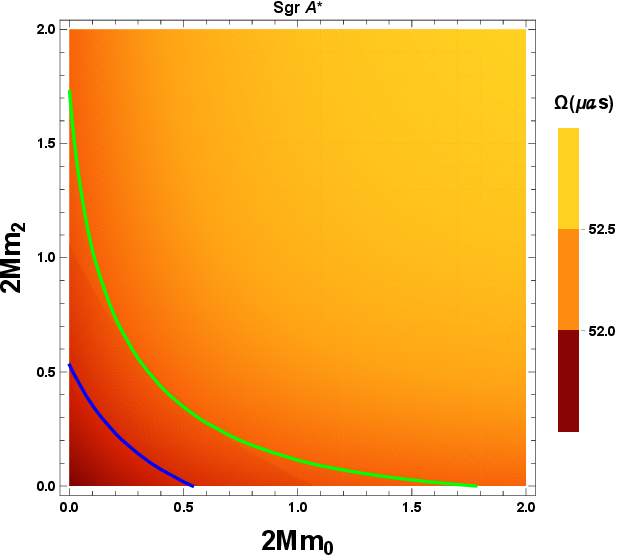}
		\end{subfigure}
		\caption{The angular diameter $\Omega(\mu as)$ of the black hole shadow correspondence to new photon sphere radius $r^L_{ph}/2M$ is presented as a function of the parameters  $S_0,~ S_2$ and  $m_0, ~m_2$ for the $M87^*$ (panels (a) and (c)) and $Sgr A^*$ (panels (b) and (d)) supermassive black holes. The enclosed regions satisfy the 1-\(\sigma\) shadow bounds for $M87^*$ and $Sgr A^*$, respectively. In panel (a), the green and blue lines correspond to $\Omega= 39.1 \, \mu as$ and $\Omega= 39.5 \, \mu as$, respectively; in panel (b), they correspond to $\Omega = 51.4 \, \mu as$ and $\Omega = 51.8 \, \mu as$, respectively; in panel (c), to $\Omega = 39.4 \, \mu as$ and $\Omega = 39.3 \, \mu as$, respectively; and in panel (d), to $\Omega = 52.2 \, \mu as$ and $\Omega = 52 \, \mu as$, respectively.}
\label{fsn2b}
\end{figure*}

The angular diameters of the black hole shadow in quadratic gravity corresponding to standard photon sphere radius $r_{ph}$ and corresponding to new photon sphere $r^L_{ph}$  obtained using the $M87^*$ and  $Sgr A^*$ data, are depicted in Fig.~\ref{fsn2} and in Fig.~\ref{fsn2b}, respectively.

\begin{table*}[htbp]
\begin{center}
\begin{tabular}{| p{2.5cm} | p{1.5cm} | p{3.cm} | p{3.cm} | p{3.cm} | p{3cm} |}
\hline
\hline
Supermassive BH  & $\Omega(\mu as)$ &$S_0/2M$ & $S_2/2M$ &$2Mm_0$ & $2Mm_2$  \\
 \hline
~&~~~ 39.1 & $0<S_0/2M\leq 0.09 $ &$0<S_2/2M\leq 0.09 $ &~~~~1 &~~~~~~1\\
$M87^*$& ~~~39.5 & $0<S_0/2M\leq 0.047 $ &$0<S_2/2M\leq 0.047 $ &~~~~1&~~~~~~1\\ 
~&~~~ 39.3 &~~~~ 0.01 &~~~~ 0.01  & $0<2Mm_0 \leq 0.39 $ &$0<2Mm_2 \leq 0.39$ \\  
~&~~~ 39.4 &~~~~ 0.01 &~~~~ 0.01  & $0<2Mm_0 \leq 0.78 $ &$0<2Mm_2 \leq 0.78$ \\  
\tableline 
~&~~~ 51.4 & $0<S_0/2M\leq 0.11 $ &$0<S_2/2M\leq 0.11$ &~~~~1 &~~~~~~1\\
$Sgr~A^*$& ~~~51.8 & $0<S_0/2M\leq 0.079 $ &$0<S_2/2M\leq 0.079 $ &~~~~1&~~~~~~1\\ 
~&~~~ 52.2 &~~~~ 0.01 &~~~~ 0.01  & $0<2Mm_0 \leq 1.8 $ &$0<2Mm_2 \leq 1.8$ \\  
~&~~~ 52 &~~~~ 0.01 &~~~~ 0.01  & $0<2Mm_0 \leq 0.55 $ &$0<2Mm_2 \leq 0.55$ \\  
\hline
\hline
\end{tabular}
\end{center}
\caption{Estimated ranges of the parameters $S_0/2M,~ S_2/2M$ and $2Mm_0, ~2Mm_2$ from the known shadow observables $\Omega (\mu as)$ for $M87^*$ and $Sgr~A^*$ in the context of standard photon sphere radius $r_{ph}$.\label{Table:1}}
\end{table*}

\begin{table*}[htbp]

\begin{tabular}{| p{2.5cm} | p{1.5cm} | p{3.cm} | p{3.cm} | p{3.cm} | p{3cm} |}
\hline
\hline
Supermassive BH  & $\Omega(\mu as)$ &$S_0/2M$ & $S_2/2M$ &$2Mm_0$ & $2Mm_2$  \\
 \hline
~&~~~ 39.1 & $0<S_0/2M\leq 0.093 $ &$0<S_2/2M\leq 0.093 $ &~~~~1 &~~~~~~1\\
$M87^*$& ~~~39.5 & $0<S_0/2M\leq 0.048 $ &$0<S_2/2M\leq 0.048 $ &~~~~1&~~~~~~1\\ 
~&~~~ 39.3 &~~~~ 0.01 &~~~~ 0.01  & $0<2Mm_0 \leq 0.38 $ &$0<2Mm_2 \leq 0.38$ \\  
~&~~~ 39.4 &~~~~ 0.01 &~~~~ 0.01  & $0<2Mm_0 \leq 0.77 $ &$0<2Mm_2 \leq 0.77$ \\  
\tableline 
~&~~~ 51.4 & $0<S_0/2M\leq 0.12 $ &$0<S_2/2M\leq 0.12$ &~~~~1 &~~~~~~1\\
$Sgr~A^*$& ~~~51.8 & $0<S_0/2M\leq 0.082 $ &$0<S_2/2M\leq 0.082 $ &~~~~1&~~~~~~1\\ 
~&~~~ 52.2 &~~~~ 0.01 &~~~~ 0.01  & $0<2Mm_0 \leq 1.74 $ &$0<2Mm_2 \leq 1.74$ \\  
~&~~~ 52 &~~~~ 0.01 &~~~~ 0.01  & $0<2Mm_0 \leq 0.54 $ &$0<2Mm_2 \leq 0.54$\\  
\hline

\hline
\end{tabular}

\caption{
 Estimated ranges of the  parameters  $S_0/2M,~ S_2/2M$ and  $2Mm_0, ~2Mm_2$ from the known shadow observables $\Omega (\mu as)$ for $M87^*$ and $Sgr~A^*$ in the context of new photon sphere radius   $r^L_{ph}$.}\label{Table:1b}
\end{table*}

\section*{Constraints from $M87^*$ }

The angular diameter of the shadow ($\Omega$) for the $M87^*$ black hole, averaged over the range from $32 \, \mu as$ to $45 \, \mu as$ with a uncertainty region $1\sigma$ from EHT observations, places constraints on the parameter pairs $(S_0, S_2)$ and $(m_0, m_2)$ of the black hole in quadratic gravity, as illustrated in Figs.~\ref{sn2a} and \ref{sn2c}.

Here, the angular diameter of the shadow ($\Omega$) is expressed as a function of the parameter pairs $(S_0, S_2)$ and $(m_0, m_2)$. In Fig.~\ref{sn2a}, the green and blue lines correspond to $\Omega = 39.1 \:\mu as$ and $\Omega = 39.5 \:\mu as$, respectively. Similarly, in Fig.~\ref{sn2c}, the green and blue lines correspond to $\Omega = 39.4 \:\, \mu as$ and $\Omega = 39.3 \:\mu as$, respectively. The blue and green curves in Figs.~\ref{sn2a} and \ref{sn2c} align with the shadow size of $M87^*$ observed by the EHT. Similar correspondence , the shadow’s angular diameter ($\Omega$) corresponding to new photon sphere radius $r^L_{ph}$, is expressed as a function of the parameter pairs $(S_0, S_2)$ and $(m_0, m_2)$ (see~in  Fig.~\ref{fsn2b}).

The constraints on the pair of parameters $(S_0, S_2)$ corresponding to the radius of the standard photon sphere $r_{ph}$ are given by (see~Table~\ref{Table:1} ):
$
0 \leq  S_0/2M \leq 0.09,
$
and $ 0 \leq  S_2/2M \leq 0.09,$
corresponding to $\Omega = 39.1 \:\mu as$. Similarly, the constraints on the parameter pair $(m_0, m_2)$ are:
$0 \leq 2M m_0 \leq 0.78,$
and
$ 0 \leq 2M m_2 \leq 0.78$, corresponding to $\Omega = 39.4 \:\mu as$. 

The constraints on the parameter pair $(S_0, S_2)$ corresponding to new photon sphere radius $r^L_{ph}$ are given by (see~Table~\ref{Table:1b} ):
$
0 \leq  S_0/2M \leq 0.093,
$
and $ 0 \leq  S_2/2M \leq 0.093,$
corresponding to $\Omega = 39.1 \:\mu as$. Similarly, the constraints on the parameter pair $(m_0, m_2)$ are:
$0 \leq 2M m_0 \leq 0.77,$
and
$ 0 \leq 2M m_2 \leq 0.77$, corresponding to $\Omega = 39.4 \:\mu as$.

Thus, $M87^*$ can be modeled within this constrained parameter space as a black hole in quadratic gravity. The numerous possible parameter points within the parameter spaces $(S_0, S_2)$ and $(m_0, m_2)$ indicate that the compatibility of the black hole model in quadratic gravity with $M87^*$ observations makes it a strong candidate for describing astrophysical black holes.

\section*{Constraints from $ {\bf Sgr ~A^*}$ }

The angular diameter of the shadow ($\Omega$) for the $Sgr A^*$ black hole, averaged over the range from $49.5 \, \mu as$ to $54.1 \, \mu as$ with an uncertainty region $1\sigma$ from EHT observations, places constraints on the parameter pairs $(S_0, S_2)$ and $(m_0, m_2)$ of the black hole in quadratic gravity, as depicted in Figs.~\ref{sn2b} and \ref{sn2d}.

Here, the angular diameter of the shadow ($\Omega$) corresponding to the radius of the standard photon sphere $r_{ph}$, is expressed as a function of the pair of parameters $(S_0, S_2)$ and $(m_0, m_2)$ (see~in Fig.~\ref{fsn2}). In Fig.~\ref{sn2b}, the green and blue lines correspond to $\Omega = 51.4 \:\mu as$ and $\Omega = 51.8 \:\mu as$, respectively. Similarly, in Fig.~\ref{sn2d}, the green and blue lines correspond to $\Omega = 52.2 \:\mu as$ and $\Omega = 52 \:\mu as$, respectively. The blue and green curves in Figs.~\ref{sn2b} and \ref{sn2d} align with the shadow size of $Sgr A^*$ observed by the EHT.  Similar correspondence , the shadow’s angular diameter ($\Omega$) corresponding to new photon sphere radius $r^L_{ph}$, is expressed as a function of the parameter pairs $(S_0, S_2)$ and $(m_0, m_2)$ (see~in Fig.~\ref{fsn2b}).

The constraints on the pair of parameters $(S_0, S_2)$ corresponding to the radius of the standard photon sphere $r_{ph}$, are given by (see Table~\ref{Table:1}):
$
0 \leq S_0/2M \leq 0.11,$
and $0 \leq S_2/2M \leq 0.11,$
corresponding to $\Omega = 51.4 \:\mu as$. Similarly, the constraints on the parameter pair $(m_0, m_2)$ are:
$0 \leq 2M m_0 \leq 1.8$ and
$0 \leq 2M m_2 \leq 1.8$, corresponding to $\Omega = 52.2 \:\mu as$.

The constraints on the parameter pair $(S_0, S_2)$ corresponding to new photon sphere radius $r^L_{ph}$, are given by (See Table~\ref{Table:1b}):
$
0 \leq S_0/2M \leq 0.12,$
and $0 \leq S_2/2M \leq 0.12,$
corresponding to $\Omega = 51.4 \:\mu as$. Similarly, the constraints on the parameter pair $(m_0, m_2)$ are:
$0 \leq 2M m_0 \leq 1.74$ and
$0 \leq 2M m_2 \leq 1.74$, corresponding to $\Omega = 52.2 \:\mu as$.

Thus, $Sgr A^*$ can be modeled as a black hole in quadratic gravity within this constrained parameter space. The numerous possible parameter points within the parameter spaces $(S_0, S_2)$ and $(m_0, m_2)$ indicate that the compatibility of the black hole model in quadratic gravity with $Sgr A^*$ observations makes it a strong candidate for describing astrophysical black holes.
\par
It is found that the parameter ranges obtained for the new photon sphere radius $r^L_{ph}$ differ slightly from those obtained for the standard photon sphere radius $r_{ph}$, although finite ranges are obtained in both cases. Thus, the Schwarzschild black hole solution in quadratic gravity emerges as a viable model.
\section{Strong gravitational lensing and its observables}\label{sect3}
In this section, we explore the phenomenon of strong gravitational lensing around a black hole in quadratic gravity. The aim is to investigate how the black hole parameters $S_0$, $S_2$, $m_0$, and $m_2$ affect the observable outcomes of strong lensing. Additionally, we compare the lensing behavior of the black hole in quadratic gravity with that of the Schwarzschild black hole ($S_0 = 0 = S_2$).

We begin our analysis by examining the strong deflection angle of a light ray in the equatorial plane ($\theta = \frac{\pi}{2}$) of the black hole.
\subsection{The strong deflection angle}
To obtain the strong deflection angle of a light ray in the equatorial plane ($\theta=\frac{\pi}{2}$) of a black hole, we rewrite first the metric (\ref{e1}) by  redefining the quantities $r$, $t$, $S_0$, $S_2$, $m_0$ and $m_2$ in units of the radius $2M$, so that $t\rightarrow t/2M$, $r\rightarrow
r/2M$, $S_0\rightarrow S_0/2M$,$S_2\rightarrow S_2/2M$,$m_0\rightarrow 2M m_0$ and $m_2\rightarrow 2M m_2$, respectively. Hence, the quadratic geometric metric (\ref{e1}) takes in the equatorial approximation the form
 \begin{equation}\label{e10}
d\bar{s}^2=-A(r)dt^2+ B(r) dr^2 +C(r) d\phi^2,
\end{equation}
where
where the metric function can be expressed as:
\begin{equation}\label{e11}
A(r ) = 1 - \frac{1}{r} + S_0 \frac{e^{-m_0 r}}{r} + S_2 \frac{e^{-m_2 r}}{r},
\end{equation}

\begin{equation}\label{e12}
B(r ) = \frac{1}{1 - \frac{1}{r} - S_0 \frac{e^{-m_0 r}}{r} (1 + m_0 r) + \frac{1}{2} S_2 \frac{e^{-m_2 r}}{r} (1 + m_2 r)}.
\end{equation}
and
\begin{equation}
  C(r)=r^2,
\end{equation}
respectively.
Gravitational lensing in the strong-field regime is governed by the interplay between the deflection angle and the lens equation. The null geodesic equation is given by
\begin{equation}\label{e14}
   \dot{r}=\frac{dr}{d\tau}= \pm\sqrt{\frac{1}{B(r)}\biggr(\frac{E^2 }{A(r)}-\frac{L^2}{r^2}\biggr)},
\end{equation}
where the constants $E$ and $L$ represent the energy and angular momentum of the particle, respectively. The parameter $\tau$ is the affine parameter along the geodesic, while the functions $A(r)$ and $B(r)$ are given by Eqs.~(\ref{e11}) and (\ref{e12}).

The unstable circular photon orbit, characterized by the radius $ r_{\text{ph}} $, can be determined by applying the conditions of the effective potential,
 given by $\left.\frac{dV_{eff}}{dr}\right|_{r_{ph}} = 0$ and $\left.\frac{d^2V_{eff}}{dr^2}\right|_{r_{ph}} < 0$. Thus, the radius of the photon sphere $r_{ph}$ is the largest real root of the equation \cite{Virbhadra:2002ju,Claudel:2000yi,Adler:2022qtb}:
\begin{equation}\label{e15}
    2A(r_{ph}) - r_{ph} A^{\prime}(r_{ph}) = 0.
\end{equation}

When a particle approaches the black hole at the closest distance $r = r_0$, where $\frac{dr}{d\tau} = 0$, the minimum impact parameter $u_0$ can be defined in terms of the closest distance $r_0$ \cite{Bozza:2002zj} as:
\begin{equation}\label{e16}
    u_0 = \frac{r_0}{\sqrt{A(r_0)}}.
\end{equation}
 
The critical impact parameter for the unstable photon orbit is expressed as:  
\begin{equation}\label{18}
    u_{ph} = \frac{r_{ph}}{\sqrt{A(r_{ph})}},
\end{equation}
where $r_{ph}$ is the radius of the photon sphere.

\begin{figure*}[htbp]
 \captionsetup[subfigure]{labelformat=simple}
    \renewcommand{\thesubfigure}{(\alph{subfigure})}
		\begin{subfigure}{.45\textwidth}
			\caption{}\label{sn3a}
			\includegraphics[height=3in, width=3in]{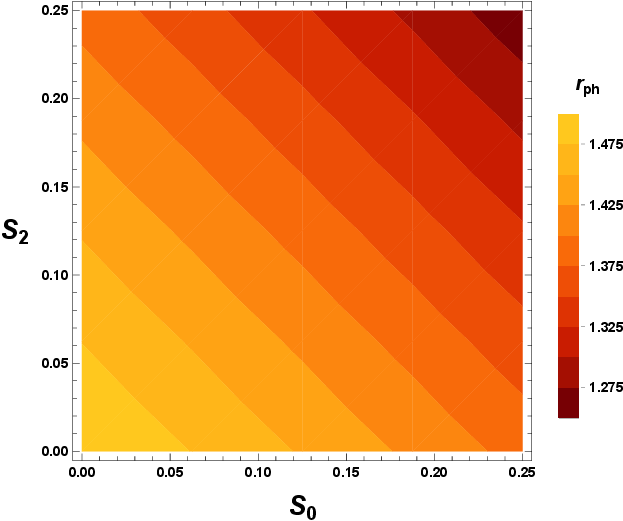}
		\end{subfigure}
            \begin{subfigure}{.4\textwidth}
			\caption{}\label{sn3b}
			\includegraphics[height=3in, width=3in]{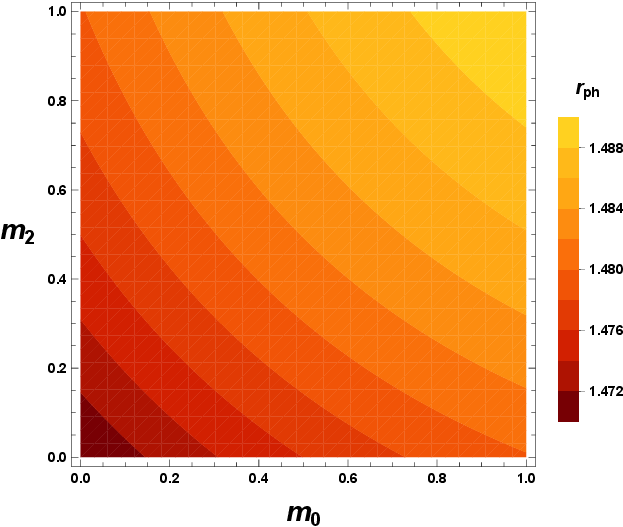}
		\end{subfigure}
		\caption{The photon sphere radius $r_{ph}$ is presented as a function of the parameters $S_0,~ S_2$ and $m_0, ~m_2$. }
		\label{fsn3}
\end{figure*}

\begin{figure*}[htbp]
 \captionsetup[subfigure]{labelformat=simple}
    \renewcommand{\thesubfigure}{(\alph{subfigure})}
		\begin{subfigure}{.45\textwidth}
			\caption{}\label{sn4a}
			\includegraphics[height=3in, width=3in]{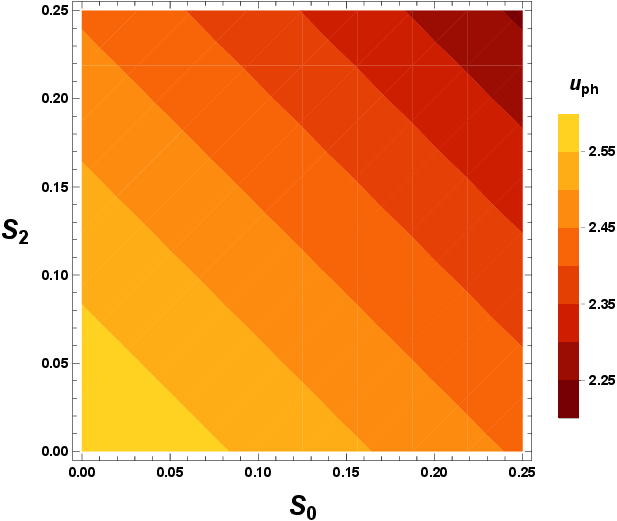}
		\end{subfigure}
            \begin{subfigure}{.4\textwidth}
			\caption{}\label{sn4b}
			\includegraphics[height=3in, width=3in]{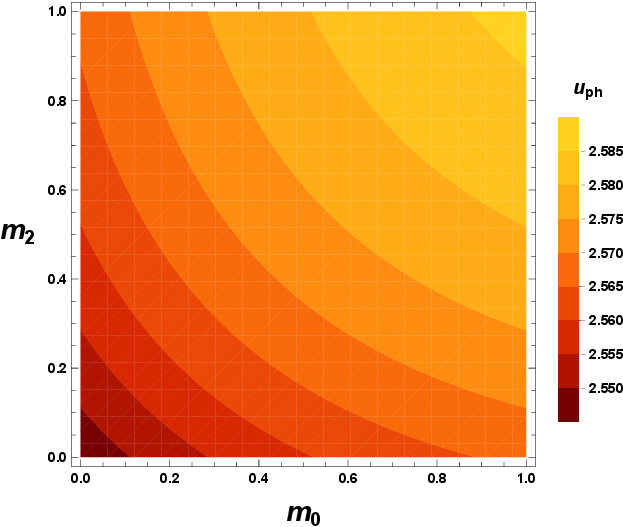}
		\end{subfigure}
		\caption{The critical impact parameter $u_{ph}$ is presented as a function of the parameters $S_0,~ S_2$ and $m_0, ~m_2$.}
		\label{fsn4}
\end{figure*}

The behavior of the photon sphere radius, $r_{ph}$ and the critical impact parameter, $u_{ph}$, are expressed as functions of the parameter pairs $(S_0, S_2)$ and $(m_0, m_2)$ (see Figs.~\ref{fsn3} and \ref{fsn4}). It is observed from Fig.~\ref{fsn3} that as the numerical values of the parameters $S_0 > 0$ or $S_2 > 0$ increase while the other parameters are kept fixed, the photon sphere radius $r_{ph}$ decreases. In contrast, an increase in the values of the parameters $m_0 > 0$ or $m_2 > 0$, with the other parameters kept constant, increases the radius of the photon sphere, $r_{ph}$. Furthermore, the radius of the photon sphere $r_{ph}$ reaches its maximum ($r_{ph}=1.5$ ) when both $S_0 \rightarrow 0$ and $S_2 \rightarrow 0$, corresponding to the Schwarzschild black hole.

It is also observed in Fig.~\ref{fsn4} that as the numerical values of the parameters $S_0 > 0$ or $S_2 > 0$ increase while the other parameters are kept fixed, the critical impact parameter, $u_{ph}$, decreases. In contrast, an increase in the values of the parameters $m_0 > 0$ or $m_2 > 0$, with the other parameters kept constant, results in an increase in the critical impact parameter, $u_{ph}$ (see Table~\ref{table:2}). Furthermore, the critical impact parameter $u_{ph}$ reaches its maximum ($u_{ph}=2.59808$ ) when both $S_0 \rightarrow 0$ and $S_2 \rightarrow 0$, corresponding to Schwarzschild black hole.

The strong deflection angle in the quadratic gravity black hole spacetime can be obtained as a function of the closest approach distance $r_0$ as \cite{Virbhadra:1998dy}
\begin{equation}\label{e18}
\alpha_D(r_0)= I(r_0)-\pi=2\int_{r_0}^\infty \frac{\sqrt{B(r)}dr}{\sqrt{C(r)} \sqrt{ \frac{A(r_0)C(r)}{A(r)C(r_0)}-1} } -\pi.
\end{equation}
The strong deflection angle $\alpha_D(r_0)$ depends on the relationship between $r_0$ and $r_{ph}$ and it increases when these two values are approximately equal (i.e., $r_0\approx r_{ph}$). Hence, we introduce a new variable $z$ defined as \cite{tsukamoto2017deflection,tsukamoto2016strong}
\begin{equation}\label{e19}
 z=1-\frac{r_0}{r}.
\end{equation}
When $r_0\approx r_{ph}$, the deflection angle takes on the following form \cite{Chagoya:2020bqz,tsukamoto2017deflection}
\begin{equation}\label{e20}
\alpha_D(u)= -\bar{a}~ log\left(\frac{u}{u_{ph}}-1\right) +\bar{b} + \mathcal{O}((u -u_{ph})log(u -u_{ph})),
\end{equation}
    where
   \begin{widetext}
    
\begin{equation}\label{e21}
\begin{split}
      & \bar{a}= \biggr[4e^{(m_0+m_2) r_{ph}} \big((r_{ph}-1) e^{(m_0+m_2) r_{ph}}+S_2 e^{m_0 r_{ph}}+S_0 e^{m_2 r_{ph}}\big)\biggr]^{\frac{1}{2}} \times \biggr[ \big(-2 S_0 \left(m_0 r_{ph}+1\right) e^{m_2 r_{ph}}+S_2 e^{m_0 r_{ph}} \left(m_2 r_{ph}+1\right)\\
      &+2 \left(r_{ph}-1\right) e^{(m_0+m_2) r_{ph}}\big) \left(m_0 S_0 \left(m_0 r_{ph}+2\right) \left(-e^{m_2 r_{ph}}\right)-m_2 S_2 e^{m_0 r_{ph}} \left(m_2 r_{ph}+2\right)+2 e^{(m_0+m_2) r_{ph}}\right)\biggr]^{-\frac{1}{2}},\\ 
\end{split}
   \end{equation}

   and
   \begin{equation}\label{e22}
      \bar{b}=b + I_R(r_{ph}),
\end{equation}
where,

\begin{equation}\label{e23}
b= \bar{a} \times \log \left(-\frac{r_{ph} \left(m_0 S_0 \left(m_0 r_{ph}+2\right) e^{m_2 r_{ph}}+m_2 S_2 e^{m_0 r_{ph}} \left(m_2 r_{ph}+2\right)-2 e^{(m_0+m_2) r_{ph}}\right)}{\left(r_{ph}-1\right) e^{(m_0+m_2) r_{ph}}+S_2 e^{m_0 r_{ph}}+S_0 e^{m_2 r_{ph}}}\right),
\end{equation}
 \end{widetext}
and
\begin{widetext}
\begin{equation}\label{e24}
I_R(r_{ph})= 2 \int_{0}^{1} r_{ph} \Bigg(\biggr[\sqrt{\frac{B(z)}{C(z)}}\biggr(\frac{A(r_{ph})}{C(r_{ph})}\frac{C(z)}{A(z)}-1\biggr)^{-\frac{1}{2}}\frac{1}{(1-z)^2} \biggr]-\frac{\bar{a}}{z ~r_{ph}}\Bigg)dz,
\end{equation}
\end{widetext}
where the above integral is computed numerically. We numerically determine the values of the critical impact parameter $u_{\text{ph}}$ and the strong lensing coefficients $\bar{a}$ and $\bar{b}$ at $r = r_{\text{ph}}$ for the quadratic gravity black hole, using the following parameter values: $S_0 = 0, 0.5$, $S_2 = 0, 0.1, 0.2, 0.3$, $m_0 = 0.5, 1$, and $m_2 = 0.3, 0.6, 0.9, 1.1$. The results are presented in Table~\ref{table:2}. 

\begin{figure*}[htbp]
 \captionsetup[subfigure]{labelformat=simple}
    \renewcommand{\thesubfigure}{(\alph{subfigure})}
		\begin{subfigure}{.45\textwidth}
			\caption{}\label{sn5a}
			\includegraphics[height=3in, width=3in]{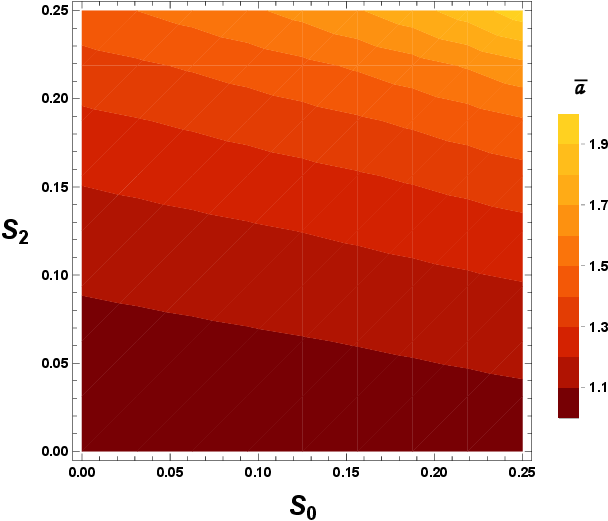}
		\end{subfigure}
            \begin{subfigure}{.4\textwidth}
			\caption{}\label{sn5b}
			\includegraphics[height=3in, width=3in]{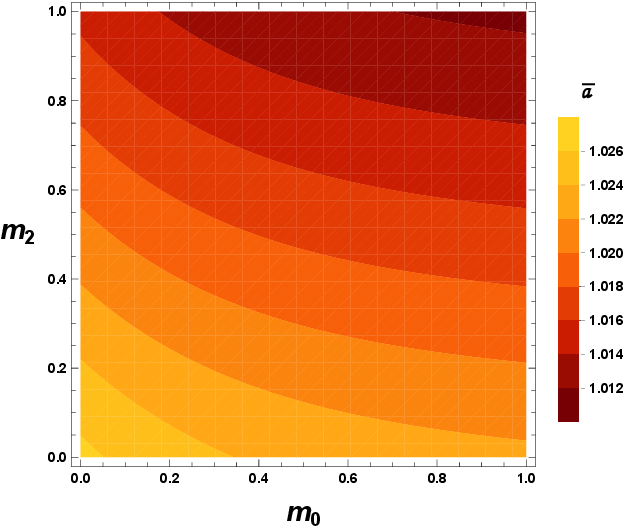}
		\end{subfigure}
		\caption{The deflection angle coefficient $\bar{a}$ is presented as a function of the parameters  $S_0,~ S_2$ and $m_0, ~m_2$.}
		\label{fsn5}
\end{figure*}

\begin{figure*}[htbp]
 \captionsetup[subfigure]{labelformat=simple}
    \renewcommand{\thesubfigure}{(\alph{subfigure})}
		\begin{subfigure}{.45\textwidth}
			\caption{}\label{sn6a}
			\includegraphics[height=3in, width=3in]{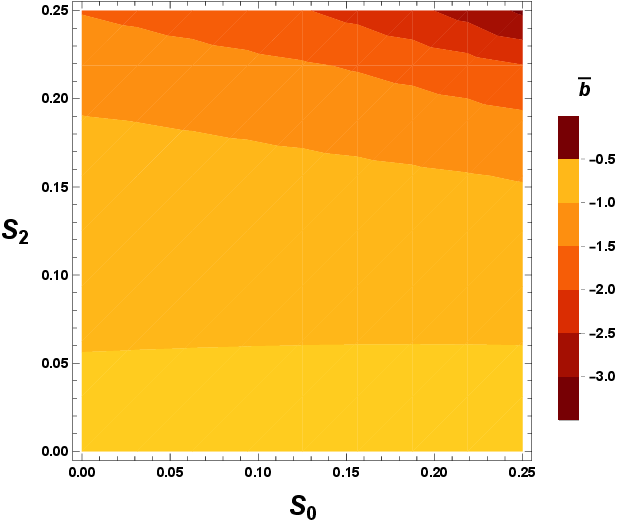}
		\end{subfigure}
            \begin{subfigure}{.4\textwidth}
			\caption{}\label{sn6b}
			\includegraphics[height=3in, width=3in]{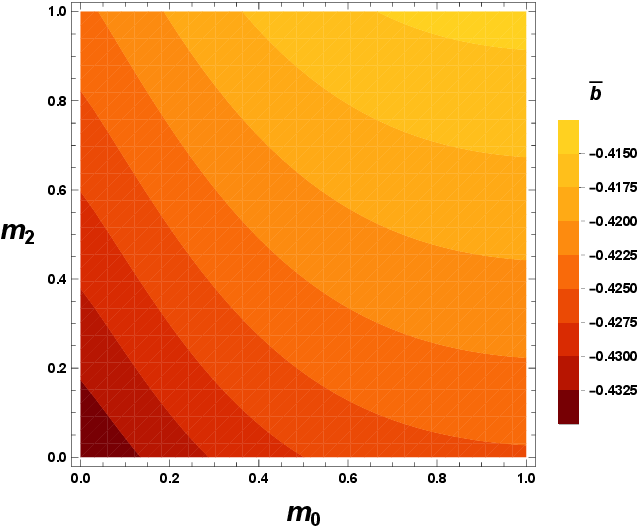}
		\end{subfigure}
		\caption{The deflection angle coefficient $\bar{b}$ is presented as a function of the parameters $S_0,~ S_2$ and $m_0, ~m_2$.}
		\label{fsn6}
\end{figure*}

The behavior of the lensing coefficients $\bar{a}$ and $\bar{b}$ is expressed as functions of the parameter pairs $(S_0, S_2)$ and $(m_0, m_2)$ (see Figs.~\ref{fsn5} and \ref{fsn6}). It is observed from Fig.~\ref{fsn5} that as the numerical values of the parameters $S_0 > 0$ or $S_2 > 0$ increase while keeping the other parameters fixed, the lensing coefficient $\bar{a}$ increases. Similarly, an increase in the values of the parameters $m_0 > 0$ or $m_2 > 0$, with the other parameters kept constant, also results in a decrease in the value of the lensing coefficient $\bar{a}$ (see Table~\ref{table:2}). The lensing coefficient $\bar{a}$ reaches the minimum value $\bar{a}=1$ \cite{Bozza:2002zj} when both $S_0 \rightarrow 0$ and $S_2 \rightarrow 0$, corresponding to the Schwarzschild black hole.
Furthermore, Fig.~\ref{fsn6} shows that as the numerical values of the parameters $S_0 > 0$ or $S_2 > 0$ increase, while the other parameters are kept fixed, the lensing coefficient $\bar{b}$ first decreases, reaches a minimum value and then increases. In contrast, an increase in the values of the parameters $m_0 > 0$ or $m_2 > 0$, with the other parameters kept constant, results in an increase in the value of the lensing coefficient $\bar{b}$ (see Table~\ref{table:2}). The lensing coefficient $\bar{b}$ reaches the value $\bar{b}=-0.40023$ \cite{Bozza:2002zj} when both $S_0 \rightarrow 0$ and $S_2 \rightarrow 0$, corresponding to the Schwarzschild black hole.

\begin{figure*}[htbp]
 \captionsetup[subfigure]{labelformat=simple}
    \renewcommand{\thesubfigure}{(\alph{subfigure})}
		\begin{subfigure}{.45\textwidth}
			\caption{}\label{sn7a}
			\includegraphics[height=3in, width=3in]{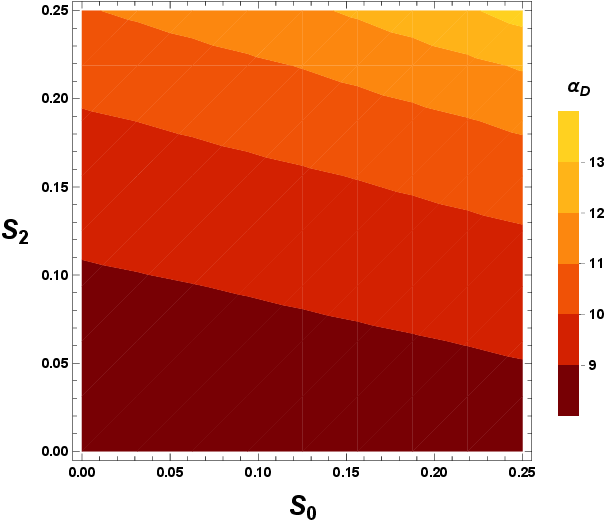}
		\end{subfigure}
            \begin{subfigure}{.4\textwidth}
			\caption{}\label{sn7b}
			\includegraphics[height=3in, width=3in]{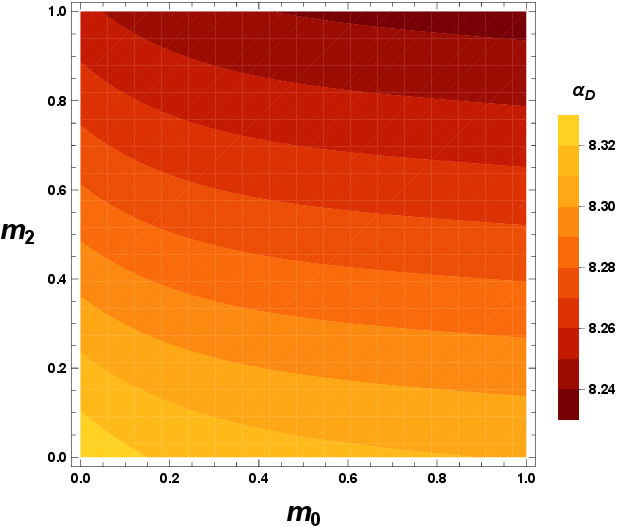}
		\end{subfigure}
  \begin{subfigure}{.52\textwidth}
			\caption{}\label{sn7c}
			\includegraphics[height=2.8in, width=2.8in]{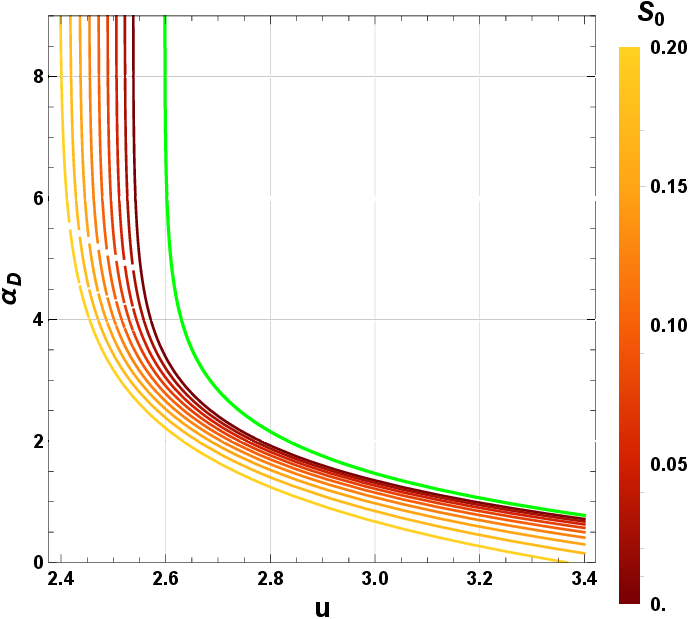}
		\end{subfigure}
            \begin{subfigure}{.4\textwidth}
			\caption{}\label{sn7d}
			\includegraphics[height=2.8in, width=2.8in]{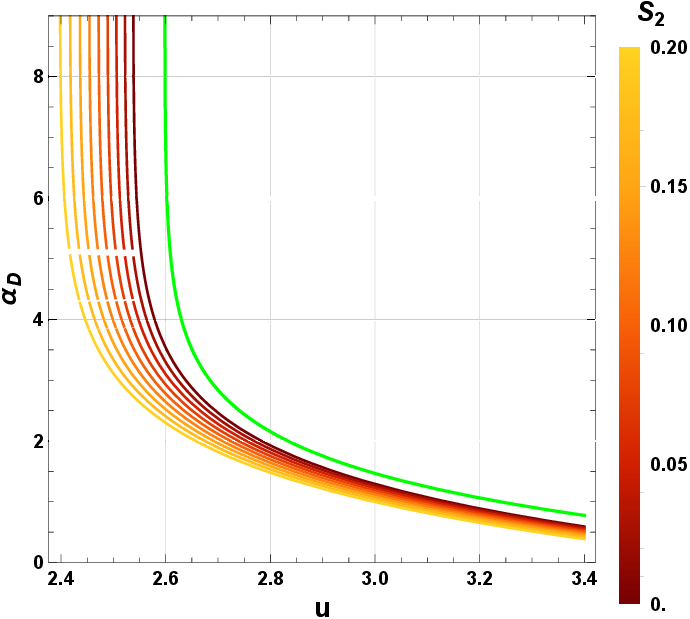}
		\end{subfigure}
		\caption{(a), (b) The strong deflection angle \(\alpha_D\) is shown as a function of the parameters \(S_0, S_2\) and \(m_0, m_2\). (c), (d) The strong deflection angle \(\alpha_D\) is presented as a function of the impact parameter \(u\), where the deflection angle diverges as \(u \rightarrow r_{ph}\). The green curve corresponds to Schwarzschild black hole ( \(S_0= S_2=0\)).}
\label{fsn7}
\end{figure*}

The behavior of the strong deflection angle, $\alpha_D$, is expressed as a function of the pair of parameters $(S_0, S_2)$ and $(m_0, m_2)$ (see Fig.~\ref{fsn7}). It is observed from Fig.~\ref{fsn7} that as the numerical values of the parameters $S_0 > 0$ or $S_2 > 0$ increase while keeping the other parameters fixed, the strong deflection angle $\alpha_D$ increases. Similarly, an increase in the values of the parameters $m_0 > 0$ or $m_2 > 0$, with the other parameters held constant, decreases the value of the strong deflection angle, $\alpha_D$. Fig.~\ref{fsn7} also shows that $\alpha_D \rightarrow
\infty $ as $u\rightarrow u_{ph}$, that is, the strong deflection angle $\alpha_D$ is unboundedly large when $u\rightarrow u_{ph}$. 
Furthermore, the strong deflection angle, $\alpha_D$, in the context of a quadratic gravity black hole is more significant than that of a Schwarzschild black hole. The quadratic gravity black hole, with the presence of parameters $S_0$, $S_2$, $m_0$ and $m_2$, can significantly intensify the gravitational lensing effect compared to other black holes. This result indicates that the gravitational lensing effect produced by a quadratic gravity black hole is significantly enhanced compared to ordinary astrophysical black holes such as the Schwarzschild black hole ($S_0=S_2=0$). Thus, the quadratic gravity black hole, influenced by moving fluids and sound waves, can be detected more quickly and distinguished from ordinary astrophysical black holes, such as the Schwarzschild black hole.


\begin{figure*}
 \captionsetup[subfigure]{labelformat=simple}
    \renewcommand{\thesubfigure}{(\alph{subfigure})}
		\begin{subfigure}{.45\textwidth}
			\caption{}\label{sn8a}
			\includegraphics[height=3in, width=3in]{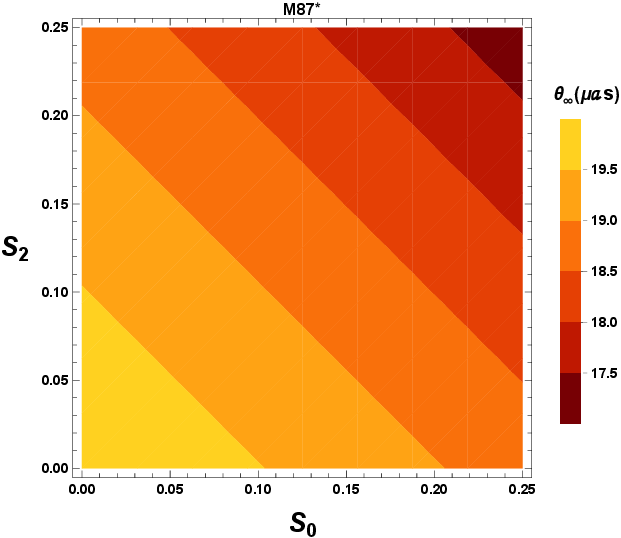}
		\end{subfigure}
            \begin{subfigure}{.4\textwidth}
			\caption{}\label{sn8b}
			\includegraphics[height=3in, width=3in]{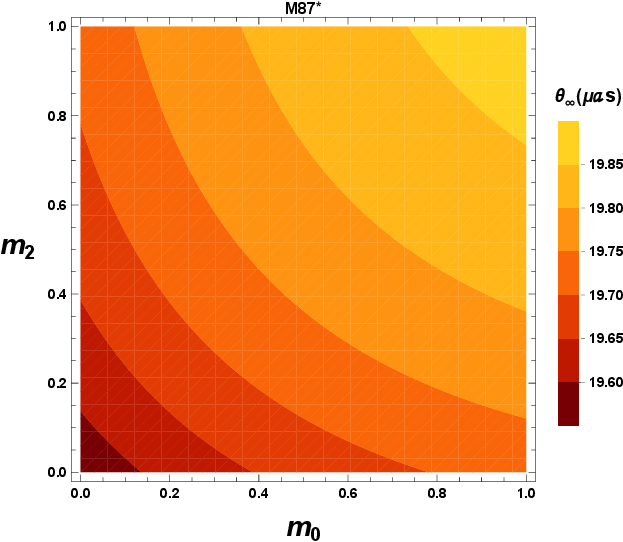}
		\end{subfigure}
		\begin{subfigure}{.45\textwidth}
			\caption{}\label{sn8c}
			\includegraphics[height=3in, width=3in]{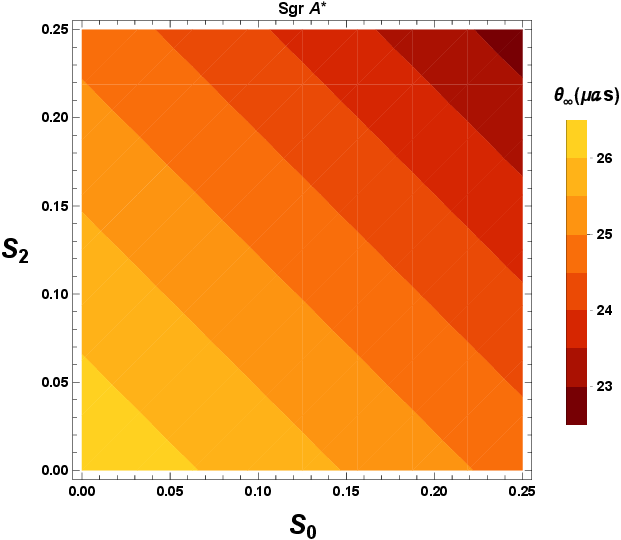}
		\end{subfigure}
            \begin{subfigure}{.4\textwidth}
			\caption{}\label{sn8d}
			\includegraphics[height=3in, width=3in]{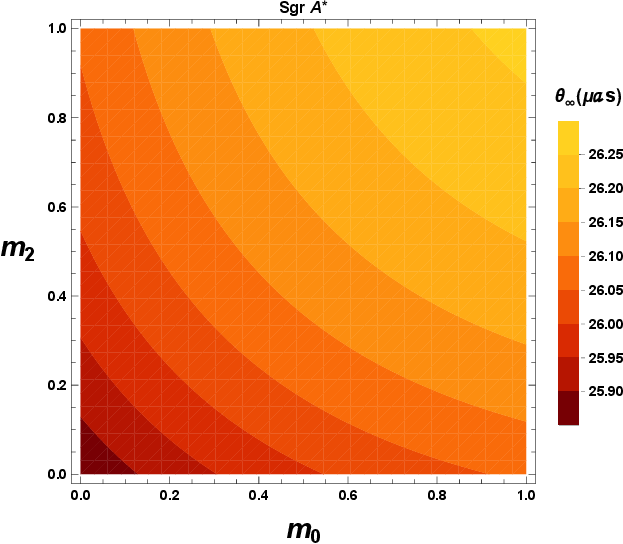}
		\end{subfigure}
		\caption{The angular position of the images $\theta_{\infty}$ ($\mu as $) is presented as a function of the parameters $S_0,~ S_2$ and $m_0, ~m_2$.}
		\label{fsn8}
\end{figure*} 
\begin{figure*}
 \captionsetup[subfigure]{labelformat=simple}
    \renewcommand{\thesubfigure}{(\alph{subfigure})}
		\begin{subfigure}{.45\textwidth}
			\caption{}\label{sn9a}
			\includegraphics[height=3in, width=3in]{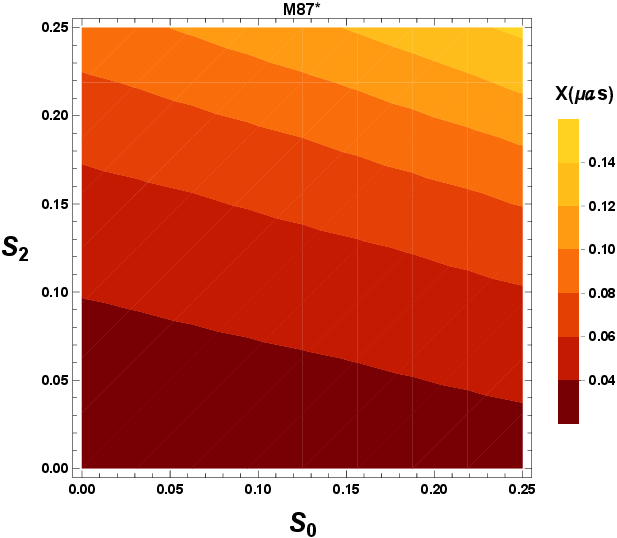}
		\end{subfigure}
            \begin{subfigure}{.4\textwidth}
			\caption{}\label{sn9b}
			\includegraphics[height=3in, width=3in]{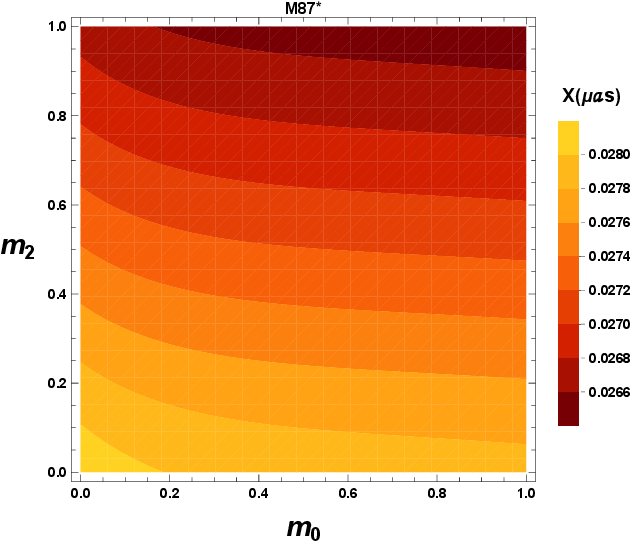}
		\end{subfigure}
		\begin{subfigure}{.45\textwidth}
			\caption{}\label{sn9c}
			\includegraphics[height=3in, width=3in]{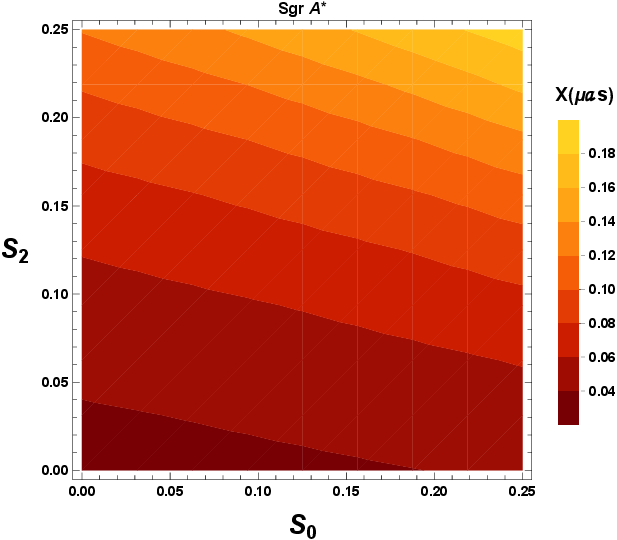}
		\end{subfigure}
            \begin{subfigure}{.4\textwidth}
			\caption{}\label{sn9d}
			\includegraphics[height=3in, width=3in]{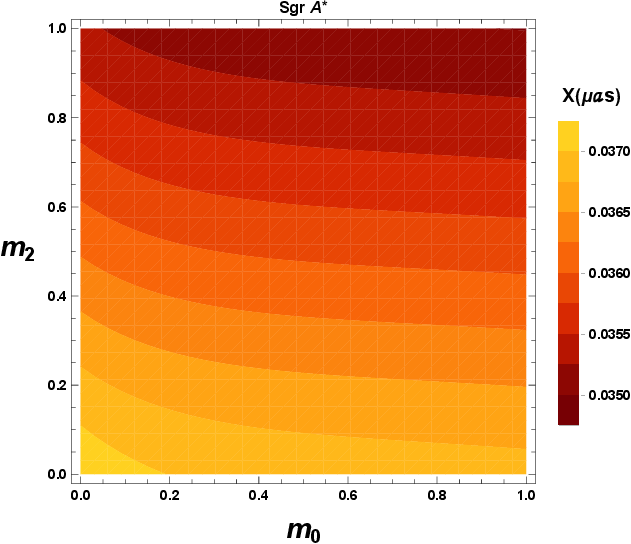}
		\end{subfigure}
		\caption{The angular separation of the images $X$($\mu as$) is presented as a function of the parameters $S_0,~ S_2$ and $m_0, ~m_2$.}
		\label{fsn9}
\end{figure*} 
\begin{figure*}
 \captionsetup[subfigure]{labelformat=simple}
    \renewcommand{\thesubfigure}{(\alph{subfigure})}
		\begin{subfigure}{.45\textwidth}
			\caption{}\label{sn10a}
			\includegraphics[height=3in, width=3in]{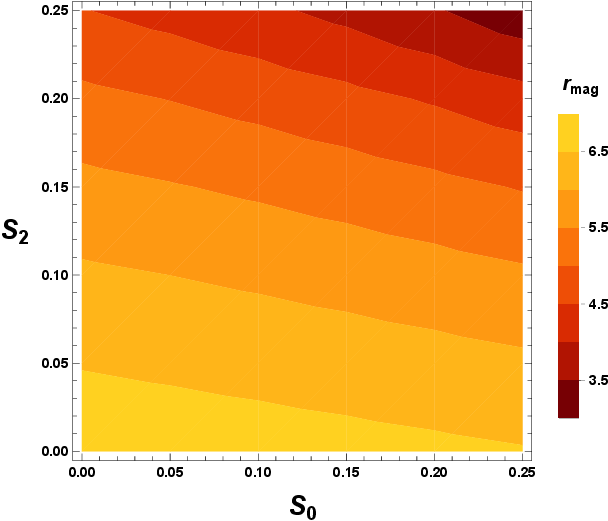}
		\end{subfigure}
            \begin{subfigure}{.4\textwidth}
			\caption{}\label{sn10b}
			\includegraphics[height=3in, width=3in]{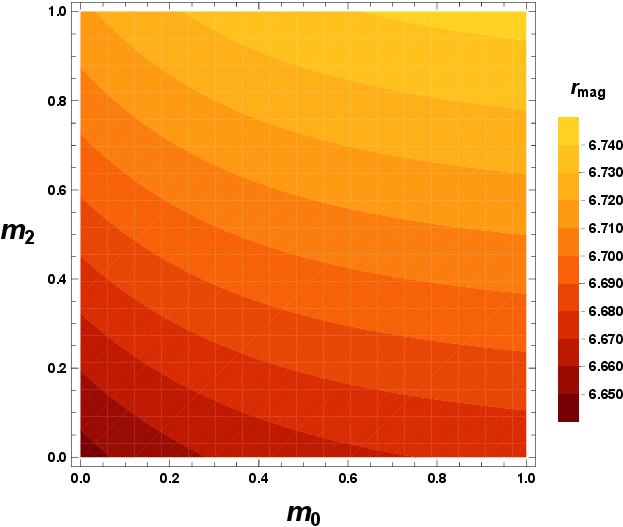}
		\end{subfigure}
		\caption{ The relative magnification of the images $r_{mag}$ is presented as a function of the parameters $S_0,~ S_2$ and $m_0, ~m_2$.}
		\label{fsn10}
\end{figure*}
\begin{table*}[htbp]
\begin{tabular}{|p{1.5cm} |p{2cm} |p{2cm} |p{2cm} |p{3cm} |p{3cm} |p{3cm}|}
\hline
\multicolumn{7}{|c|}{Strong Lensing Coefficients}\\
\hline
$S_0$ & $m_0$ & $S_2$ & $m_2$ & $\bar{a}$  & $\bar{b} $ & $u_{ph}$ \\
\hline
 $0$ & $ $ & $0$  & $ $ & 1& $ - 0.40023$ & $2.59808$ \\
\hline
 &  &  & $0.3$ & 1.31945 & -1.00257 &2.28715\\
 &  & $0.1$ & $0.6$ &1.27941 & -0.882235& 2.34857\\
 &  &  & $0.9$ & 1.25713 & -0.821543 & 2.39032\\
 &  &  & 1.1&1.24752&-0.800193 &2.41006\\
 &  &  &  & & &\\
 &  &  & && &\\
 &  &  & $0.3$ & 1.52133& -1.5402 & 2.09035\\
 & $0.5$ & $0.2$ & $0.6$ & 1.3881 & -1.0993 & 2.21375\\
 &  &  & $0.9$ & 1.32338& -0.910493 &2.30134\\
 &  &  & 1.1&1.29772&-0.849069 &2.34341\\
 &  &  & && &\\
  &  &  &  & & &\\
 &  &  & $0.3$ & 2.03915 &-3.44553 &1.87746\\
 &  & $0.3$ & $0.6$ & 1.58237& -1.60264& 2.06229\\
 &  &  & $0.9$ &1.42174 &-1.08157 &2.20068\\
 &  &  & 1.1&1.36647& -0.936446&2.26848\\
 &  &  & && &\\
$0.1$ &  &  &  & & &\\
&  &  & && &\\
 &  &  & $0.3$ &1.20153 & -0.753994 &2.35682\\
 &  & $0.1$ & $0.6$ &1.17352 &-0.675696 &2.41898\\
 &  &  & $0.9$ &1.15807 &-0.638187 &2.46065\\
 &  &  & 1.1&1.15139& -0.626371&2.48013\\
 &  &  & && &\\
 &  &  &  & & &\\
 &  &  & $0.3$ & 1.33508&-1.04248 &2.16208\\
 & $1$ & $0.2$ & $0.6$ &1.24902 & -0.78693&2.28751\\
 &  &  & $0.9$ & 1.20606& -0.676661&2.37513\\
 &  &  & 1.1&1.18869& -0.643378&2.41666\\
 &  &  &&& &\\
  &  &  &  & & &\\
 &  &  & $0.3$ &1.61347 &-1.84353 &1.95072\\
 &  & $0.3$ & $0.6$ &1.37237 & -1.02851 &2.13973\\
 &  &  & $0.9$ & 1.27383 &-0.753786 &2.27858\\
 &  &  & 1.1&1.23801&-0.678277&2.34553\\
 
\hline
\end{tabular}
\caption{Estimation of the strong lensing coefficients for different values of the quadratic gravity black hole parameters $S_0=0,0.1$, $S_2=0,0.1,0.2,0.3$, $ m_0=0.5,1$ and $ m_2=0.3,0.6,0.9,1.1$ respectively.\label{table:2}}
\end{table*}

\subsection{Lensing observables}
We now turn to exploring the observable features of strong gravitational lensing within the framework of quadratic gravity black hole geometry. In this analysis, we assume that both the observer and the source are positioned at large distances from the black hole (acting as the lens) with near-perfect alignment. Furthermore, the source is assumed to lie directly behind the black hole. Under these conditions, the lens equation is expressed as
\cite{Bozza:2001xd}
\begin{equation}\label{e25}
 \beta=\theta-\frac{D_{ls}}{D_{os}}\Delta \alpha_{n}.
\end{equation}
where $\Delta \alpha_{n} = \alpha_D(\theta) - 2n\pi$ denotes the offset deflection angle, with $\alpha_D(\theta)$ representing the primary deflection angle, and $n$ indicating the number of loops or revolutions a photon completes around the black hole. The angles $\beta$ and $\theta$ correspond to the angular separations between the lens (black hole) and the source and between the image and the source, respectively. The distances $D_{ol}$, $D_{ls}$ and $D_{os}$ refer to the observer-lens, lens-source and observer-source distances, respectively, with the relation $D_{os} = D_{ol} + D_{ls}$.

From Eqs.~\eqref{e20} and \eqref{e25}, the angular separation between the black hole (lens) and the $n^{\text{th}}$ relativistic image is given by
\begin{equation}\label{e26}
     \theta _n =  \theta^0 _n - \frac{u_{ph}e_n(\theta_n^0-\beta)D_{os}}{\bar{a}D_{ol}D_{ls}}.
 \end{equation}
Here, $\theta^0_n$ denotes the angular position of the image corresponding to a photon completing $2n\pi$ rotations around the black hole (lens). In the framework of strong gravitational lensing, where surface brightness remains conserved, the magnification of the relativistic effect is defined as the ratio of the solid angle subtended by the $n^{\text{th}}$ image to that subtended by the source \cite{Virbhadra:1999nm}.

The magnification for the $n^{\text{th}}$ relativistic image is given by \cite{Bozza:2002zj}
\begin{equation}\label{e27}
\mu_n=\biggr(\frac{\beta}{\theta}\frac{d\beta}{d\theta}\biggr)^{-1}\biggr|_{\theta_n^0}=\frac{ u^2_{ph}(1+e_n)e_n D_{os}}{\beta~ \bar{a}D_{ls}D^2_{ol}}.
\end{equation}
The above equation indicates that the first relativistic image exhibits the highest brightness, with the magnification decreasing exponentially as $n$ increases. In other words, the brightness of the first image exceeds that of the subsequent relativistic images. Moreover, Eq.~\eqref{e27} becomes unbounded as $\beta \rightarrow 0$, implying that perfect alignment improves the chances of detecting relativistic images.

We focus on the scenario where the brightest image, i.e., the outermost image at $\theta_1$, is resolved as a distinct image, while the remaining inner images cluster near $\theta_{\infty}$, defined as $\theta_n |{n \rightarrow \infty}:= \theta{\infty}$. Using the deflection angle from Eq.~\eqref{e21}, one can derive key strong lensing observables, including the angular position of the image cluster at $\theta_{\infty}$, the angular separation $S$ between the outermost and innermost images, and the relative magnification $r_{mag}$ between the outermost and inner relativistic images. These quantities are given as \cite{Bozza:2002zj, Kumar:2022fqo}:
\begin{equation}\label{e28}
\theta_{\infty}=\frac{u_{ph}}{D_{ol}},
\end{equation}
 \begin{equation}\label{e29}
X= \theta_1-\theta_{\infty}\approx\theta_{\infty}e^\frac{(\bar{b} -2\pi)}{\bar{a}},
  \end{equation}
\begin{equation}\label{e30}
  r_{mag}=\frac{\mu_1}{\Sigma^\infty_{n=2}\mu_{n}}\approx \frac{5\pi}{\bar{a}~log(10)}.
\end{equation}
If the strong lensing observables $\theta_{\infty}$, $X$ and $r_{mag}$ are determined from observations, the lensing coefficients $\bar{a}$, $\bar{b}$ and the minimum impact parameter $u_{ph}$ can be computed by inverting Eqs.~\eqref{e28}, \eqref{e29} and \eqref{e30}, respectively. Additionally, the observed data can be compared with the corresponding theoretical predictions. These results enable the identification and distinction of the quadratic gravity black hole from a Schwarzschild black hole.

\begin{table*}[htbp]
\begin{tabular}{|p{1cm}|p{1cm}|p{1cm}|p{1cm}|p{2.5cm}|p{2.5cm}|p{2.5cm}|p{2.5cm}|p{2.5cm}|}
        \hline
        \multicolumn{4}{|c|}{\textbf{Parameters}} & \multicolumn{2}{|c|}{\textbf{$ M87^*$}} & \multicolumn{2}{|c|}{\textbf{$ Sgr A^*$}} & \multicolumn{1}{|c|}{\textbf{$ M87^*$, $ Sgr A^*$}} \\
        \hline
        $S_0$ & $m_0$ & $S_2$ & $m_2$ & $ \theta_{\infty} (\mu as)$ & $X(\mu as)$ & $ \theta_{\infty} (\mu as)$ & $X(\mu as)$  & $r_{mag}$ \\

\hline
\hline
0&&0&&19.9632&0.0249839&26.3826&0.0330177&6.82188\\
\hline
&&&0.3&17.5741&0.0702727&23.2252&0.0928695&5.17023\\
&&0.1&0.6&18.0461&0.0666955&23.849&0.0881421&5.33203\\
&&&0.9&18.3668&0.0645049&24.2729&0.085247&5.42657\\
&&&1.1&18.5185&0.0633437&24.4733&0.0837124&5.46834\\
&&&&&&&&\\
&&&&&&&&\\
&&&0.3&16.0619&0.0938527&21.2268&0.124032&4.48417\\
&0.5&0.2&0.6&17.0101&0.0833594&22.4799&0.110164&4.91454\\
&&&0.9&17.6831&0.0770561&23.3693&0.101834&5.15489\\
&&&1.1&18.0064&0.0738842&23.7965&0.0976423&5.25682\\
&&&&&&&&\\
&&&&&&&&\\
&&&0.3&14.4262&0.12222&19.065&0.161522&3.34546\\
&&0.3&0.6&15.8463&0.108542&20.9418&0.143445&4.31118\\
&&&0.9&16.9097&0.0951575&22.3472&0.125756&4.79827\\
&&&1.1&17.4306&0.0884627&23.0356&0.116909&4.99233\\
&&&&&&&&\\
0.1&&&&&&&&\\
&&&&&&&&\\
&&&0.3&18.1094&0.0517972&23.9327&0.068453&5.67767\\
&&0.1&0.6&18.5871&0.0494169&24.5639&0.0653073&5.81319\\
&&&0.9&18.9073&0.0479746&24.987&0.0634013&5.89072\\
&&&1.1&19.0569&0.0471875&25.1849&0.0623611&5.92491\\
&&&&&&&&\\
&&&&&&&&\\
&&&0.3&16.6131&0.068777&21.9552&0.0908929&5.10973\\
&1&0.2&0.6&17.5769&0.0611796&23.2289&0.0808525&5.46178\\
&&&0.9&18.2502&0.0568936&24.1187&0.0751882&5.65635\\
&&&1.1&18.5693&0.0547179&24.5404&0.072313&5.73901\\
&&&&&&&&\\
&&&&&&&&\\
&&&0.3&14.9891&0.0973464&19.8089&0.128649&4.22809\\
&&0.3&0.6&16.4413&0.079819&21.7282&0.105486&4.97088\\
&&&0.9&17.5083&0.0698379&23.1383&0.0922948&5.3554\\
&&&1.1&18.0227&0.065123&23.8181&0.0860639&5.51037\\
\hline
\end{tabular}
\caption{Estimation of the strong lensing observables for the Supermassive Black Holes $ M87^*$, $ Sgr A^*$ for different values of the quadratic gravity black hole parameters $S_0=0.1$, $m_0=0.5, 1$, $m_2=0.3,0.6,0.9,1.1$  and $S_2=0.1,0.2,0.3$ respectively. The observable quantity $r_{mag}$ does not depend on the mass or the distance of the black hole from the observer.}\label{table:3}
\end{table*}

\subsubsection{Lensing observables and the quadratic gravity black hole parameters}
Considering the supermassive black holes $ M87^*$ and
$ Sgr A^*$
in the center of the nearby galaxies, we numerically  obtain the observable quantities
$\theta_{\infty}$ ,$X$ and $r_{mag}$ for the  quadratic gravity black hole  (see Table~\ref{table:3}).
The mass and distance from the earth for $ M87^*$ \cite{EventHorizonTelescope:2019dse} are  $M\approx 6.5 \times 10^9 M_{\odot}
$,$D_{ol}\approx 16.8Mpc$, for $ Sgr A^*$ are  $M\approx
4.28\times 10^6M_{\odot}$,$D_{ol}\approx 8.32kpc$
\cite{gillessen2017update} . 
The behavior of the observable quantities
$\theta_{\infty}$ ,$X$ and $r_{mag}$  for the supermassive black holes $ M87^*$ and
$Sgr A^*$ are expressed as functions of the parameter pairs $(S_0, S_2)$ and $(m_0, m_2)$ (see Figs.~\ref{fsn8},\ref{fsn9} and \ref{fsn10}). It is observed from Fig.~\ref{fsn8} that as the numerical values of the parameters $S_0 > 0$ or $S_2 > 0$ increase while keeping the other parameters fixed, the observable quantity
$\theta_{\infty}$ decreases. Similarly, an increase in the values of the parameters $m_0 > 0$ or $m_2 > 0$, with the other parameters held constant, also increases the value of the observable quantity $\theta_{\infty}$ (see Table~\ref{table:3}). The observable quantity
$\theta_{\infty}$ reaches the maximum value when both $S_0 \rightarrow 0$ and $S_2 \rightarrow 0$, corresponding to Schwarzschild black hole.
Furthermore, Fig.~\ref{fsn9} shows that as the numerical values of the parameters $S_0 > 0$ or $S_2 > 0$ increase, while keeping the other parameters fixed, $X$ increases. Conversely, an increase in the values of the parameters $m_0 > 0$ or $m_2 > 0$, with the other parameters held constant, results in a decrease in the value of the observable quantity $X$ (see Table~\ref{table:3}). The observable quantity
$X$ reaches the minimum value when both $S_0 \rightarrow 0$ and $S_2 \rightarrow 0$, corresponding to Schwarzschild black hole.

It is also observed from Fig.~\ref{fsn10} that as the numerical values of the parameters $S_0 > 0$ or $S_2 > 0$ increase while keeping the other parameters fixed, the observable quantity
$r_{mag}$ decreases. Similarly, an increase in the values of the parameters $m_0 > 0$ or $m_2 > 0$, with the other parameters held constant, also increases the value of the observable quantity $r_{mag}$ (see Table~\ref{table:3}). The observable quantity
$r_{mag}$ reaches the maximum value when $S_0 \rightarrow 0$ and $S_2 \rightarrow 0$ correspond to Schwarzschild's black hole.

 It is also observed that the relative magnification $r_{mag}$ of the relativistic images remains independent of the black hole's mass and distance.
\subsection{Einstein rings}
When the black hole (lens), the source, and the observer are perfectly aligned, corresponding to \( \beta = 0 \), the black hole bends light in every direction, creating a ring-like image.
 This phenomenon, known as the Einstein ring, has been extensively studied in \cite{Einstein:1936llh, Liebes:1964zz, Mellier:1998pk, Bartelmann:1999yn, Schmidt:2008hc, Guzik:2009cm}.
By simplifying Eq.~\eqref{e26} for $\beta = 0$, the angular radius of the $n^{\text{th}}$ relativistic image can be expressed as
\begin{equation}\label{e31}
     \theta _n =  \theta^0 _n \biggr(1 - \frac{u_{ph} e_n  D_{os}}{\bar{a}D_{ls}D_{ol}}\biggr).
 \end{equation}
In the case where the black hole (lens) is situated halfway between the source and the observer, with \( D_{os} = 2D_{ol} \), and assuming that \( D_{ol} \gg u_{ph} \), the angular radius of the $n^{\text{th}}$ relativistic Einstein ring within the spacetime of a quadratic gravity black hole can be expressed as follows
\begin{equation}\label{31}
\theta^E_n=\frac{ u_{ph}(1+e_n)}{D_{ol}}.
\end{equation}

\begin{figure*}[htbp]
 \captionsetup[subfigure]{labelformat=simple}
    \renewcommand{\thesubfigure}{(\alph{subfigure})}
		\begin{subfigure}{.45\textwidth}
			\caption{}\label{sn11a}
			\includegraphics[height=3in, width=3in]{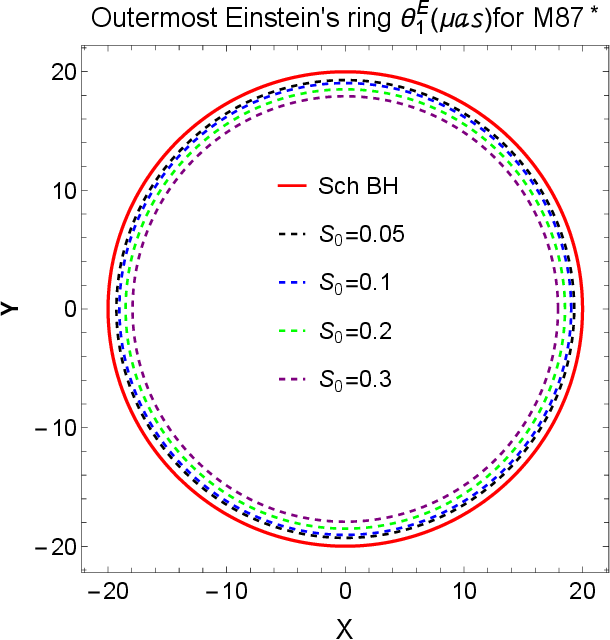}
		\end{subfigure}
            \begin{subfigure}{.4\textwidth}
			\caption{}\label{sn11b}
			\includegraphics[height=3in, width=3in]{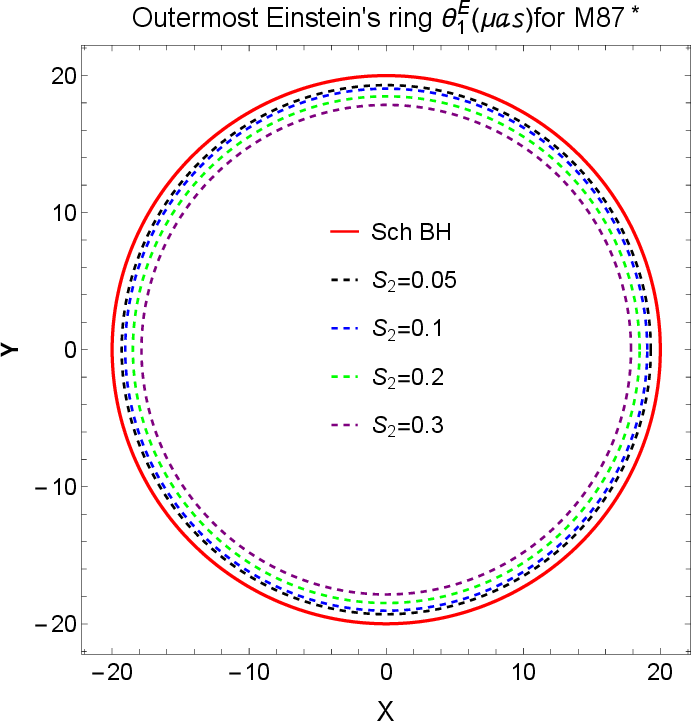}
		\end{subfigure}
  \begin{subfigure}{.45\textwidth}
			\caption{}\label{sn11c}
			\includegraphics[height=3in, width=3in]{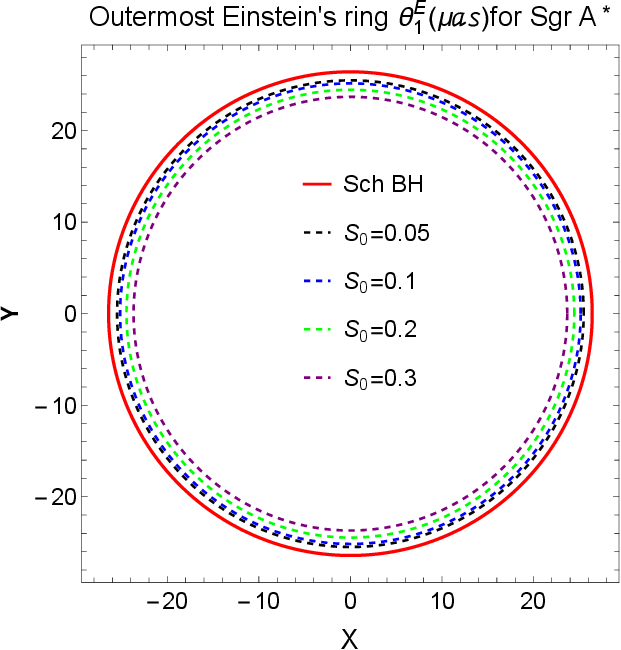}
		\end{subfigure}
            \begin{subfigure}{.4\textwidth}
			\caption{}\label{sn11d}
			\includegraphics[height=3in, width=3in]{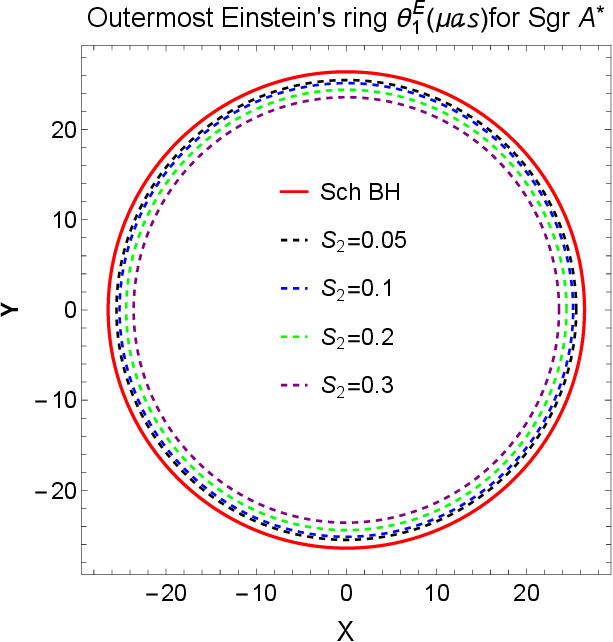}
		\end{subfigure}
		\caption{The outermost Einstein ring $\theta^E_1$ ($\mu as$) of the black hole in quadratic gravity, in the context of $M87^*$  and $Sgr A^*$, is presented in the X-Y plane. The black hole parameters $S_0$ and $S_2$ are varied independently, while the other parameters remain fixed. }
		\label{fsn11}
\end{figure*} 

The angular radius $\theta^E_1$ represents the outermost Einstein ring, as shown in Fig.~\ref{fsn11} for supermassive black holes $M87^*$ and $Sgr A^*$. From Fig.~\ref{fsn11}, it is observed that the angular radius $\theta^E_1$ of the outermost Einstein ring decreases with increasing value of the parameter $S_0$ or parameter $S_2$.
From Fig.~\ref{fsn11}, it is also evident that the angular radius $\theta^E_1$ of the outermost Einstein ring for the quadratic gravity black hole is smaller than the corresponding angular radius of the Schwarzschild black hole.
\subsection{Time delay in the strong field limit}
\begin{figure*}[htbp]
 \captionsetup[subfigure]{labelformat=simple}
    \renewcommand{\thesubfigure}{(\alph{subfigure})}
		\begin{subfigure}{.45\textwidth}
			\caption{}\label{sn8a}
			\includegraphics[height=3in, width=3in]{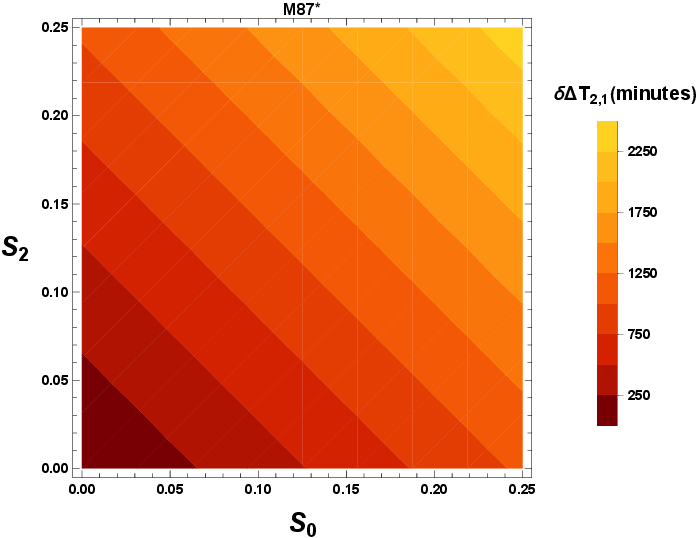}
		\end{subfigure}
            \begin{subfigure}{.4\textwidth}
			\caption{}\label{sn8b}
			\includegraphics[height=3in, width=3in]{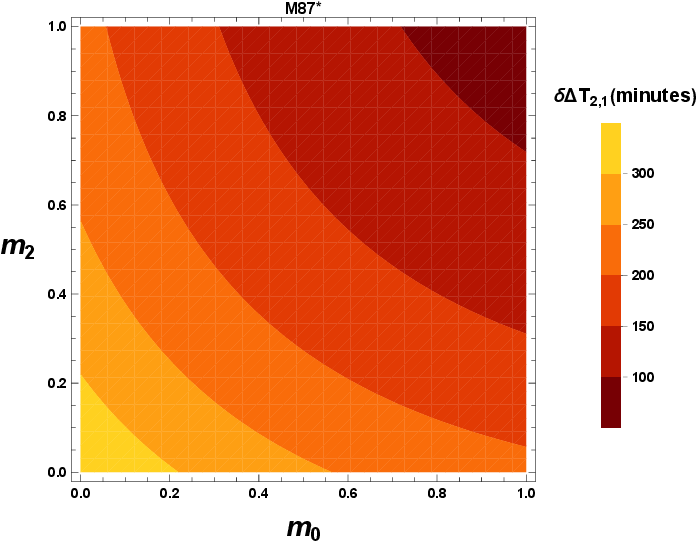}
		\end{subfigure}
		\begin{subfigure}{.45\textwidth}
			\caption{}\label{sn8c}
			\includegraphics[height=3in, width=3in]{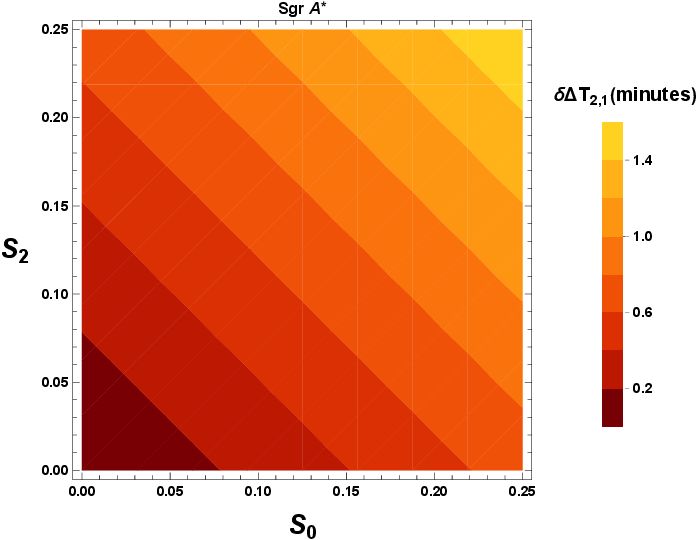}
		\end{subfigure}
            \begin{subfigure}{.4\textwidth}
			\caption{}\label{sn8d}
			\includegraphics[height=3in, width=3in]{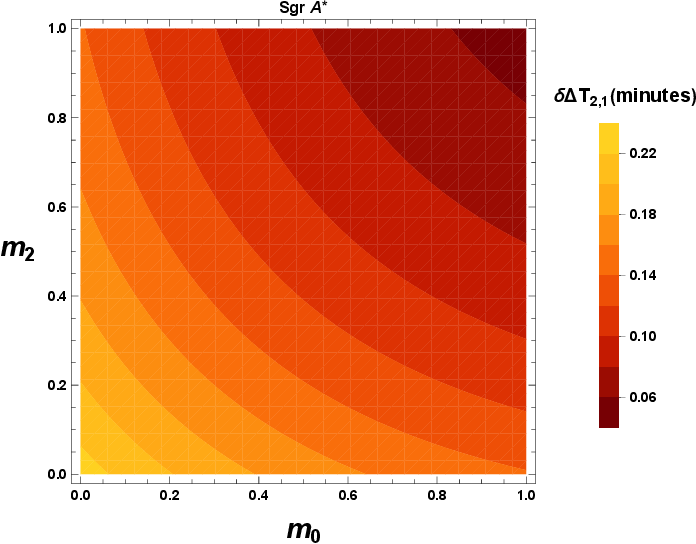}
		\end{subfigure}
		\caption{The deviation of the time delays between the first and second relativistic
images, for the quadratic gravity black hole from Schwarzschild black hole  $\delta \Delta T_{2,1}= \Delta T_{2,1}^{Sch}-\Delta T_{2,1}^{QG}$ minutes is presented as a function of the parameters  $S_0,~ S_2$ and  $m_0, ~m_2$.}
		\label{fsn12}
\end{figure*}

Time delay is one of the most significant observables in the strong gravitational lensing effect. This arises from the time difference between the formation of two relativistic images. This time difference occurs because the photon traverses different paths around the black hole. Since the travel time for photon paths associated with distinct relativistic images is different, a time difference exists between these images. If the time signals from two relativistic images are obtained through observations, one can calculate the time delay between the two signals \cite{Virbhadra:2007kw,Bozza:2003cp}. The time taken by a photon to orbit the black hole is given by \cite{Bozza:2003cp}
 \begin{equation}\label{e33}
\tilde{T}=\tilde{a}\log\biggr(\frac{u}{u_{ph}}-1\biggr)+\tilde{b}+\mathcal{O}(u-u_{ph}).
\end{equation}
Using the above Eq.~(\ref{e33}), one can calculate the time difference between two relativistic images. In the context of a spherically symmetric static black hole spacetime, the time delay between two relativistic images, when both images are on the same side of the black hole, is expressed as follows
\begin{equation}\label{33}
\Delta T_{2,1}=2\pi u_{ph}=2\pi D_{ol} \theta_{\infty}.
\end{equation}
Using the above equation, the time delays for a Schwarzschild black hole ($S_0=S_2=0$) and the quadratic gravity black hole, with $S_0=S_2=0.1$ and $m_0=m_2=1$ have been estimated in Table~\ref{table:4} for several supermassive black holes located in the centers of nearby galaxies. The deviation of the time delays between the first and second relativistic
images, for the quadratic gravity black hole from the Schwarzschild black hole $\delta \Delta T_{2,1}= \Delta T_{2,1}^{Sch}-\Delta T_{2,1}^{ASquadratic gravity}$minutes with $S_0=S_2=0.1$ and $m_0=m_2=1$ has been displayed for the cases of supermassive black holes $M87^{*}$ and $Sgr A^{*}$ in Fig.~\ref{fsn12}.
 Fig.~\ref{fsn12} shows that as the numerical values of the parameters $S_0 > 0$ or $S_2 > 0$ increase while keeping the other parameters fixed, $\delta \Delta T_{2,1}$ increases. Conversely, an increase in the values of the parameters $m_0 > 0$ or $m_2 > 0$, with the other parameters held constant, decreases the value of the observable quantity $\delta \Delta T_{2,1}$.

\begin{table*}
\begin{tabular}{| p{3.0cm}|p{3.cm}|p{3.cm}|p{3.5cm}|p{3.5cm}|p{3.0cm}|}
\hline
 Galaxy  &$M(M_{\odot})$& $D_{ol}(Mpc)$&
$\Delta T_{2,1}$ Schwarzschild\vfill BH ($S_0=S_2=0$) &  $\Delta T_{2,1}$ quadratic gravity BH\vfill($S_0=S_2=0.1$,\vfill$m_0=m_2=1$)  \\
\hline
NGC 4395 & $3.6\times 10^5 $& 4.3&0.962522&0.915472\\
Milky Way &  4.3 $\times$ $10^6$ & 0.0083&11.4968&10.9348\\
NGC 7457 & $8.95\times 10^6 $& 12.53&23.9294&22.7596\\
NGC 4486A & $1.44\times 10^7 $& 18.36&38.5009&36.6189\\
NGC 2778 & $1.45\times 10^7 $&23.44&38.7682&36.8732\\
NGC 3607 & $1.37\times 10^8 $& 22.65&366.293&348.388\\
NGC 4026 & $1.80\times 10^8 $& 13.35&481.261&457.736\\
NGC 5576 & $2.73\times 10^8 $& 25.68&729.912&694.233\\
NGC 7052 & $3.96\times 10^8 $& 70.4&1058.77&1007.02\\
NGC 3379 & $4.16\times 10^8 $& 10.70&1112.25&1057.88\\
NGC 4261 & $5.29\times 10^8 $& 32.36&1414.37&1345.23\\
NGC 6251 & $6.14\times 10^8 $& 108.4&1641.63&1561.39\\
NGC 5077 & $8.55\times 10^8 $& 38.7&2285.99&2174.25\\
NGC 7768 & $1.34\times 10^9 $& 116.0&3582.72&3407.59\\
NGC 6861 & $2.10\times 10^9 $& 28.71&5614.71&5340.25\\
NGC 4751 & $2.44\times 10^9 $& 32.81&6523.76&6204.86\\
Cygnus A & $2.66\times 10^9 $& 242.7&7111.95&6764.32\\
NGC 5516 & $3.69\times 10^9 $& 55.3&9865.85&9383.59\\
NGC 4649 & $4.72\times 10^9 $& 16.46&12619.7&12002.9\\
M87  & $6.5\times 10^9 $& 16.68&17378.9&16529.4\\
NGC 3842 & $9.09\times 10^9 $& 92.2&24303.7&23115.7\\
\hline
\end{tabular}
\caption{Estimation of the time delay for various supermassive BHs for the black hole-like spacetime in quadratic gravity and the Schwarzschild spacetime, respectively. The masses and distances are given in solar masses and megaparsecs (Mpc), respectively \cite{Kormendy:2013dxa}. The time delays $\Delta T_{2,1}$ are estimated in minutes.
\label{table:4}}
\end{table*}

\subsection{Results and discussions}
Using observations of strong gravitational fields, we employed the methodology introduced by Bozza \cite{Bozza:2002zj} to differentiate between various types of spherically symmetric black holes and explore their astrophysical consequences, especially for supermassive black holes at the centers of nearby galaxies. This approach has been widely used to test the fundamental properties of black holes in quadratic gravity, comparing them with the standard Schwarzschild solution in general relativity. In our analysis, we examined several key aspects of astrophysical significance, including the calculation of strong lensing coefficients $\bar{a}$, $\bar{b}$ and $u_{ph}$ (see Table~\ref{table:2}). Using these coefficients, we derived the strong deflection angle along with several observable quantities, such as the angular position of the innermost image $\theta_{\infty}$, the image separation $X$ and the magnification ratio $r_{mag}$, specifically for the supermassive black holes $M87^*$ and $Sgr A^*$ (see Table~\ref{table:3}). Furthermore, we examined the time delay between relativistic images (see Table~\ref{table:4}) for multiple supermassive black holes located at the centers of nearby galaxies.
\\\\
\paragraph{\textbf{Angular position $\theta _{\infty }$ for the set of images:}} We have numerically determined the angular positions of the relativistic images for various supermassive black holes within the quadratic gravity framework. The results for supermassive black holes $M87^*$ and $Sgr A^*$ are summarized in Table~\ref{new1}.
\begin{table}[htbp]
\begin{center}
\begin{tabular}{|c|c|c|c|c|c|}
\hline
SMBH &$S_0$ &$m_{0}$& $S_{2}$ & $m_{2}$ & $\theta _{\infty }$ ($\mu as$) \\
\hline
$M87^{\ast }$&0.1&0.5& 0.1 & $0.3\leq m_{2}\leq1.1$ & $\theta _{\infty }\in
(17.57,18.52)$ \\
\hline
$Sgr A^{\ast} $& 0.1&0.5&  0.1 & $0.3\leq m_{2}\leq1.1$ & $\theta _{\infty }\in(23.22,24.48)$ \\
\hline
$M87^{\ast }$&0.1&0.5& 0.2 & $0.3\leq m_{2}\leq1.1$ & $\theta _{\infty }\in
(16.06,18.01)$ \\
\hline
$Sgr A^{\ast} $& 0.1&0.5&  0.2 & $0.3\leq m_{2}\leq1.1$ & $\theta _{\infty }\in(21.22,23.80)$ \\
\hline
$M87^{\ast }$&0.1&0.5& 0.3 & $0.3\leq m_{2}\leq1.1$ & $\theta _{\infty }\in
(14.42,17.44)$ \\
\hline
$Sgr A^{\ast} $& 0.1&0.5&  0.3 & $0.3\leq m_{2}\leq1.1$ & $\theta _{\infty }\in(19.06,23.04)$ \\
\hline
$M87^{\ast }$&0.1&1& 0.1 & $0.3\leq m_{2}\leq1.1$ & $\theta _{\infty }\in
(18.10,19.06)$ \\
\hline
$Sgr A^{\ast} $& 0.1&1&  0.1 & $0.3\leq m_{2}\leq1.1$ & $\theta _{\infty }\in(23.93,25.19)$ \\
\hline
$M87^{\ast }$&0.1&1& 0.2 & $0.3\leq m_{2}\leq1.1$ & $\theta _{\infty }\in
(16.61,18.57)$ \\
\hline
$Sgr A^{\ast} $& 0.1&1&  0.2 & $0.3\leq m_{2}\leq1.1$ & $\theta _{\infty }\in(21.95,24.55)$ \\
\hline
$M87^{\ast }$&0.1&1& 0.3 & $0.3\leq m_{2}\leq1.1$ & $\theta _{\infty }\in
(14.98,18.03)$ \\
\hline
$Sgr A^{\ast} $& 0.1&1&  0.3 & $0.3\leq m_{2}\leq1.1$ & $\theta _{\infty }\in(19.80,23.82)$ \\
\hline
\end{tabular}
\end{center}
\caption{Admissible range for the angular position of the innermost relativistic images $\theta_{\infty}$, is analyzed for the supermassive black holes $M87^*$ and $Sgr A^*$. This range is determined for varying values of the quadratic gravity black hole parameters $S_2$, $m_0$ and $m_2$ while keeping the parameter $S_0$ constant.}\label{new1}
\end{table}
As one can see from Tables~\ref{table:2} and \ref{new1}, the angular position of the innermost relativistic images $\theta_{\infty}$ depends on the values of the quadratic gravity black hole parameters $S_2$, $m_0$ and $m_2$. An increase in the value of $m_2$ from 0.3 to 1.1 will lead, for the black holes $M87^{*}$ and $Sgr A^{*}$, to an increase in $\theta_{\infty}$ for fixed values of the remaining parameters, while $\theta_{\infty}$ decreases with an increase in the parameter $S_2$ from 0.1 to 0.3, assuming the other parameters are fixed. Furthermore, considering the same mass and distance of the black hole, it is observed that the angular position of the innermost relativistic images $\theta_{\infty}$ for the quadratic gravity black hole is smaller than that in the case of a Schwarzschild black hole. Consequently, fundamental quadratic gravity has a significant effect on the potentially observable angular positions of the images $\theta_{\infty}$, allowing a distinction between these images and those associated with other classical black holes.\\\\
\paragraph{\textbf{The angular separation:}}We have numerically obtained the angular separation between the outermost and innermost relativistic images $X$ for the supermassive black holes $M87^*$ and $Sgr A^*$ within the framework of quadratic gravity, as summarized in Table~\ref{new2}.
\begin{table}[htbp]
\begin{center}
\begin{tabular}{|c|c|c|c|c|c|}
\hline
SMBH& $S_0$&$m_{0}$ & $S_{2}$ & $m_{2}$ & $X$ ($\mu as$) \\
\hline
$M87^{\ast }$&0.1&0.5& 0.1 & $0.3\leq m_{2}\leq1.1$ & $X\in
(0.06,0.08)$ \\
\hline
$Sgr A^{\ast} $&0.1&0.5&  0.1 & $0.3\leq m_{2}\leq1.1$ & $X\in
(0.08,0.10)$ \\
\hline
$M87^{\ast }$&0.1&0.5& 0.2 & $0.3\leq m_{2}\leq1.1$ & $X\in
(0.07,0.10)$ \\
\hline
$Sgr A^{\ast} $&0.1&0.5&  0.2 & $0.3\leq m_{2}\leq1.1$ & $X\in
(0.09,0.13)$ \\
\hline
$M87^{\ast }$&0.1&0.5& 0.3 & $0.3\leq m_{2}\leq1.1$ & $X\in
(0.08,0.13)$ \\
\hline
$Sgr A^{\ast} $&0.1&0.5&  0.3 & $0.3\leq m_{2}\leq1.1$ & $X\in
(0.11,0.17)$ \\
\hline
$M87^{\ast }$&0.1&1& 0.1 & $0.3\leq m_{2}\leq1.1$ & $X\in
(0.04,0.06)$ \\
\hline
$Sgr A^{\ast} $&0.1&1&  0.1 & $0.3\leq m_{2}\leq1.1$ & $X\in
(0.06,0.07)$ \\
\hline
$M87^{\ast }$&0.1&1& 0.2 & $0.3\leq m_{2}\leq1.1$ & $X\in
(0.05,0.07)$ \\
\hline
$Sgr A^{\ast} $&0.1&1&  0.2 & $0.3\leq m_{2}\leq1.1$ & $X\in
(0.07,0.10)$ \\
\hline
$M87^{\ast }$&0.1&1& 0.3 & $0.3\leq m_{2}\leq1.1$ & $X\in
(0.06,0.10)$ \\
\hline
$Sgr A^{\ast} $&0.1&1&  0.3 & $0.3\leq m_{2}\leq1.1$ & $X\in
(0.08,0.13)$ \\
\hline
\end{tabular}
\end{center}
\caption{Admissible range for the angular separation between outermost and innermost relativistic images $X$, is analyzed for the supermassive black holes $M87^*$ and $Sgr A^*$. This range is determined for varying values of the quadratic gravity parameters $S_2$, $m_0$ and $m_2$ while keeping the parameter $S_0$ constant.}\label{new2}
\end{table}\\\\
As one can see from Tables~\ref{table:2} and \ref{new2}, the angular separation between the outermost and innermost relativistic images $X$ depends on the values of the quadratic gravity black hole parameters $S_2$, $m_0$ and $m_2$. An increase in the value of $m_2$ from 0.3 to 1.1 will lead, for the black holes $M87^{*}$ and $Sgr A^{*}$, to a decrease in $X$ for fixed values of the remaining parameters, while $X$ increases with an increase in the parameter $S_2$ from 0.1 to 0.3, assuming the other parameters are fixed. Furthermore, considering the same mass and distance of the black hole, it is observed that the angular separation between outermost and innermost relativistic images $X$ for the quadratic gravity black hole is larger than that in the case of a Schwarzschild black hole. Consequently, fundamental quadratic gravity has a significant effect on the potentially observable angular separation between the outermost and innermost relativistic images $X$, allowing a distinction between these images and those associated with other classical black holes.

\paragraph{\textbf{Relative magnification:}} 
We have numerically obtained the relative magnification of the relativistic images $r_{mag}$for the supermassive black holes $M87^*$ and $Sgr A^*$ within the framework of quadratic gravity, as summarized in Table~\ref{new3}.
\begin{table}[htbp]
\begin{center}
\begin{tabular}{|c|c|c|c|c|c|}
\hline
SMBH& $S_0$&$m_{0}$ & $S_{2}$ & $m_{2}$ &  $r _{mag}$ magnitude \\
\hline
$M87^{\ast }$ ,$Sgr A^{\ast} $&0.1&0.5& 0.1 & $0.3\leq m_{2}\leq1.1$ &  $r _{mag}\in
(5.17,5.47)$ \\
\hline
$M87^{\ast }$ ,$Sgr A^{\ast} $&0.1&0.5& 0.2 & $0.3\leq m_{2}\leq1.1$ &  $r _{mag}\in
(4.48,5.26)$ \\
\hline
$M87^{\ast }$ ,$Sgr A^{\ast} $&0.1&0.5& 0.3 & $0.3\leq m_{2}\leq1.1$ &  $r _{mag}\in
(3.34,5.00)$ \\
\hline
$M87^{\ast }$ ,$Sgr A^{\ast} $&0.1&1& 0.1 & $0.3\leq m_{2}\leq1.1$ &  $r _{mag}\in
(5.67,5.93)$ \\
\hline
$M87^{\ast }$ ,$Sgr A^{\ast} $&0.1&1& 0.2 & $0.3\leq m_{2}\leq1.1$ &  $r _{mag}\in
(5.10,5.74)$ \\
\hline
$M87^{\ast }$ ,$Sgr A^{\ast} $&0.1&1& 0.3 & $0.3\leq m_{2}\leq1.1$ &  $r _{mag}\in
(4.22,5.52)$ \\
\hline

\hline
\end{tabular}
\end{center}
\caption{Admissible range for the relative magnification of the relativistic images $r_{mag}$, is analyzed for the supermassive black holes $M87^*$ and $Sgr A^*$. This range is determined for varying values of the quadratic gravity parameters $S_2$, $m_0$ and $m_2$ while keeping the parameter $S_0$ constant. }\label{new3}
\end{table}

As illustrated in Tables~\ref{table:2} and \ref{new3}, the relative magnification of the relativistic images, denoted as $r_{mag}$, is influenced by the parameters of the quadratic gravity black hole, specifically $S_2$, $m_0$ and $m_2$. Increasing the value of $m_2$ from 0.3 to 1.1 results in a corresponding rise in $r_{mag}$ for the supermassive black holes $M87^{*}$ and $Sgr A^{*}$, assuming the other parameters remain constant. Conversely, $r_{mag}$ decreases as the parameter $S_2$ increases from 0.1 to 0.3, again with other parameters held fixed.

Moreover, when comparing black holes of the same mass and distance, it is found that the relative magnification $r_{mag}$ for the quadratic gravity black hole is less than that for a Schwarzschild black hole. This indicates that the fundamental aspects of quadratic gravity significantly affect the observable relative magnification of the relativistic images, allowing for differentiation between these images and those linked to other classical black holes. It is also important to note that the relative magnification of the relativistic images is independent of the black hole's mass or distance.
\\\\

\paragraph{\textbf{The Einstein rings:}} 
For the supermassive black holes $M87^{*}$ and $Sgr A^{*}$, the outermost Einstein rings $\theta_1^E$ are presented in Figs.~\ref{sn11a} and \ref{sn11c} for the parameter values $m_0 = m_2 = 1$, $S_0 = 0.1$ and $S_2 = 0, 0.05, 0.1, 0.2, \;\& \; 0.3$. Furthermore, the cases with $m_0 = m_2 = 1$, $S_2 = 0.1$, and $S_0 = 0, 0.05, 0.1, 0.2, \;\& \; 0.3$ are shown in Figs.~\ref{sn11b} and \ref{sn11d}.
Our findings indicate that the radius of the outermost Einstein ring, $\theta_1^E$, decreases as the values of the quadratic gravity black hole parameters $S_0$ or $S_2$ increase while keeping the remaining parameters fixed. Notably, the Einstein ring $\theta_1^E$ for a quadratic gravity black hole is significantly smaller than that of the standard classical Schwarzschild black hole ($S_0 = S_2 = 0$).
\\\\
\paragraph{\textbf{Relativistic time delay.}}
We have also evaluated the time delays $\Delta T_{2,1}$ between the first and second-order relativistic images for various supermassive black holes within the framework of the quadratic gravity black hole, using the parameter values $(S_0 = S_2 = 0.1, \, m_0 = m_2 = 1)$, as well as for the Schwarzschild black hole $(S_0 = S_2 = 0)$. The time delay $\Delta T_{2,1}$ for the quadratic gravity black hole (e.g., $\sim 23115.7$ minutes for NGC $3842$) is significantly smaller compared to the Schwarzschild black hole (e.g., $\sim 24303.7$ minutes for NGC $3842$). 
Thus, the deviation in the time delays for the quadratic gravity black hole with $(S_0 = S_2 = 0.1, \, m_0 = m_2 = 1)$ from the standard Schwarzschild black hole $(S_0 = S_2 = 0)$ is $
\delta \Delta T_{2,1} = \Delta T_{2,1}^{Sch} - \Delta T_{2,1}^{QG} = 1188 , \text{minutes}.
$
This suggests that if the first and second relativistic images are distinguishable, the time delay between them could serve as a valuable indicator to differentiate the quadratic gravity black hole from the classical Schwarzschild black hole.

\section{Conclusions and final remarks}\label{sect4}
This paper explores the shadow and strong gravitational lensing phenomena associated with black holes in quadratic gravity. These black holes are described by a solution derived from the quantum effective action and the corresponding quantum equations of motion, which are obtained from multi-graviton correlation functions in quadratic gravity \cite{Pawlowski:2023dda}.
 A black hole in quadratic gravity is described by five parameters, with $M$ representing the mass of the black hole. The two free parameters $S_0$ and $S_2$ characterize the strength of the exponentially decaying Yukawa corrections. Additionally, the masses $m_0$ and $m_2$ of the spin-0 and spin-2 modes are associated with various couplings of curvature terms. This has been discussed in detail in \cite{Pawlowski:2023dda}.

Equations~\eqref{e1}-\eqref{e3} provide the parametrization of asymptotically flat solutions within the linearized regime. These equations ensure that the asymptotically flat solutions extend smoothly to \( r \to \infty \). By systematically varying the parameters, one can identify a variety of topologically distinct solutions, each characterized by specific scaling behaviors near \( r = 0 \) or, in some cases, by a termination point \( r_\text{term} > 0 \). The defining features of these solution classes (Type~Ia, Type~Ib, Type~Ic, Type~II and Type~III) emerge from the distinct initial conditions outlined in Table~2 of the referenced literature \cite{daas2023probing}.
From the perspective of quadratic gravity, the geometry described by Eqs.~\eqref{e1}-\eqref{e3} represents the gravitational field in the linear approximation. However, this framework breaks down in the strong gravity regime. In this regime, the event horizon becomes sensitive to exponentially small Yukawa corrections and non-zero values of $S_0$ and $S_2$ can give rise to wormholes and naked singularities. Consequently, the standard formulas used for shadow imaging must be applied with caution, as new photon trajectories can emerge that would otherwise be blocked by an event horizon. A detailed analysis of this phenomenon is available in the literature \cite{daas2023probing}. Quadratic gravity introduces significant effects on Event Horizon Telescope (EHT) observations, with profound implications across multiple domains. The Schwarzschild-like geometry described by Eqs.~\eqref{e1}-\eqref{e3} is discovered to be part of a diverse solution space that includes horizonless configurations, naked singularities and wormhole structures. The inclusion of Yukawa corrections and additional parameters like \( S_0 \) and \( S_2 \) opens up new photon trajectories that do not occur in standard black hole models. Consequently, traditional shadow imaging methods must be carefully reconsidered as these new pathways alter the observed shadow properties. A significant observation is the impact of such modifications on the photon ring and shadow size, particularly for TypeIc and TypeII geometries, which exhibit distinguishable features in their intensity profiles (see \cite{daas2023probing} for more details). These findings emphasize that shadow observations using the EHT can effectively constrain the parameter space of quadratic gravity.
Their findings lead to two significant implications. First, the EHT demonstrates the ability to explore and even exclude certain regions of the phase space associated with static, spherically symmetric and asymptotically flat vacuum solutions in quadratic gravity. Second, this phase space includes horizonless naked singularities and wormhole geometries with intensity profiles nearly identical to those produced by the Schwarzschild geometry. Consequently, based on these observations alone, it remains challenging to determine whether the imaged object genuinely has an event horizon.
In the present paper, we have constrained the black hole parameters $S_0$, $S_2$, $m_0$, and $m_2$ through shadow observables (the angular diameter of the black hole shadow) using the EHT observation data for the supermassive black holes M87* and Sgr A*. This was done for both cases: the new photon sphere radius $r^L_{ph}$ and the standard photon sphere radius $r_{ph}$. It is observed that, in both cases, the obtained results fall within a finite parameter space and are consistent with the EHT observation data for M87* and Sgr A*. Our work interprets the EHT observational results for black-hole-type solutions within the framework of a clear quantum gravity perspective, grounded in observational data. Thus, the Schwarzschild black hole solution in quadratic gravity emerges as a viable model.

The classical Schwarzschild black hole is obtained when $S_0 =S_2=0$. The present study highlights that
traditional black hole solutions derived from classical
general relativity may not fully capture the complexities of black holes in the context of quantum gravity.
The asymptotic safety scenario suggests that gravitational interactions approach a fixed point at high-energy scales, indicating new physics beyond classical theories. Quantum black hole solutions in quadratic gravity exhibit significant
deviations from classical predictions, particularly outside the event horizon. These deviations could manifest in gravitational lensing phenomena, where the curvature of spacetime influences the paths of light rays passing near massive objects.
In this study, we primarily restrict our analysis to positive values of the black hole parameters $S_0$,$S_2$,$m_0$ and $m_2$.
The study of light deflection (gravitational lensing) in the strong field limit, particularly in regions dominated by the gravitational effects of quadratic gravity black holes, offers powerful theoretical and observational tools for distinguishing between standard general relativistic models and modified gravity models. Although the radial coordinate dependence of the quadratic gravity metric takes a fixed form, when interpreted as a galactic metric, the bending angle of light and other optical effects are significantly influenced by the metric parameters, as well as by the baryonic mass and radius of the supermassive black hole. The quadratic gravity metric dictates the specific values of the bending angles and other lensing effects, and these values differ substantially from those predicted by the classical Schwarzschild metric.

For example, when $S_0=0.1$,$S_2=0.1$,$m_0=1$, and $m_2=1$, the time delay of the photons is much smaller in quadratic gravity than in standard general relativity. This may be related to the gravitational light deflection angle being larger than the value predicted by the standard general relativistic approach. We observe significant differences in the lensing effects when comparing our results for the quadratic gravity black holes with those of standard Schwarzschild black holes.
  
 Moreover, the study of black hole-like solutions in quadratic gravity can be distinguished from standard astrophysical black holes, such as the Schwarzschild black holes($S_0=0$,$S_2=0$), through shadow and strong gravitational lensing observations.
  
  As a result, the quadratic gravity black hole, influenced by moving fluids and sound waves, can be detected more easily and distinguished from ordinary astrophysical black holes, such as the Schwarzschild black hole. Consequently, the study of gravitational lensing may provide evidence for the existence of quadratic gravity effects in the Universe. Our findings represent a promising and significant step toward connecting the fundamental theory of quadratic gravity to potentially feasible black hole spacetimes.
\\\\
\section*{Acknowledgements}
 FR would like to thank the authorities of the Inter-University Centre for Astronomy and Astrophysics, Pune, India for providing research facilities. This work is a part of the project submitted in DST-SERB, Govt. of India. N.U.M would like to thank CSIR, Govt. of India for providing Senior Research Fellowship (No. 08/003(0141)/2020-EMR-I). G. Mustafa is very thankful to Prof. Gao Xianlong from the Department of Physics, Zhejiang Normal University, for his kind support and
help during this research.

\bibliographystyle{apsrev4-1}%
\bibliography{rm.bib} 


\end{document}